\documentclass[prb,preprint]{revtex4-2} 

\usepackage{float}
\usepackage{amsmath} 
\usepackage{amsfonts} 
\usepackage{graphicx} 
\usepackage{braket}
\usepackage{verbatim}
\usepackage{hyperref}

\begin{document}


\title{A Step-by-Step Superconducting Quantum Chip Design Tutorial: from Theory to Simulation}

\author{Hiu Yung Wong}
\email{hiuyung.wong@sjsu.edu} 
\author{James Saslow}
\author{Shreyan Juvvadi}

\affiliation{Department of Electrical Engineering, San Jose State University, San Jose, CA 95192}



\date{\today}

\begin{abstract}
This paper provides a step-by-step tutorial on superconducting qubit chip design. It covers the design of the coplanar waveguide, feedline, resonator, flux-bias line, transmon qubit, and the interactions among these components. Finally, the design of a 3-qubit chip with two coupled qubits is demonstrated. Each design is accompanied by basic physics and engineering theory, followed by simulations and calculations using Ansys High Frequency Structure Simulator (HFSS), Q3D Extractor, Maxwell 3D Design, and/or Quantum Metal. YouTube videos are produced to demonstrate the setup and simulation procedures for each section, and the Quantum Metal source code and HFSS projects are available on GitHub. Each section contains suggested exercises for students to practice the theory and skills learned. The pyEPR package, run through Quantum Metal, is used to calculate transmon-resonator cross-Kerr and transmon anharmonicity. After the tutorial, the students will have a better understanding of the concepts and calculations of the scattering matrix, quality factor, qubit lifetime, magnetic flux quantum, cross-Kerr, and anharmonicity. They will also master the skills and know-how to perform eigenmode, scattering-matrix, magnetostatic, and energy-participation-ratio (EPR) simulations and calculations.

\end{abstract}

\maketitle 

\section{Introduction} 

\begin{figure}[H]
\centering
\includegraphics[width=1\textwidth]{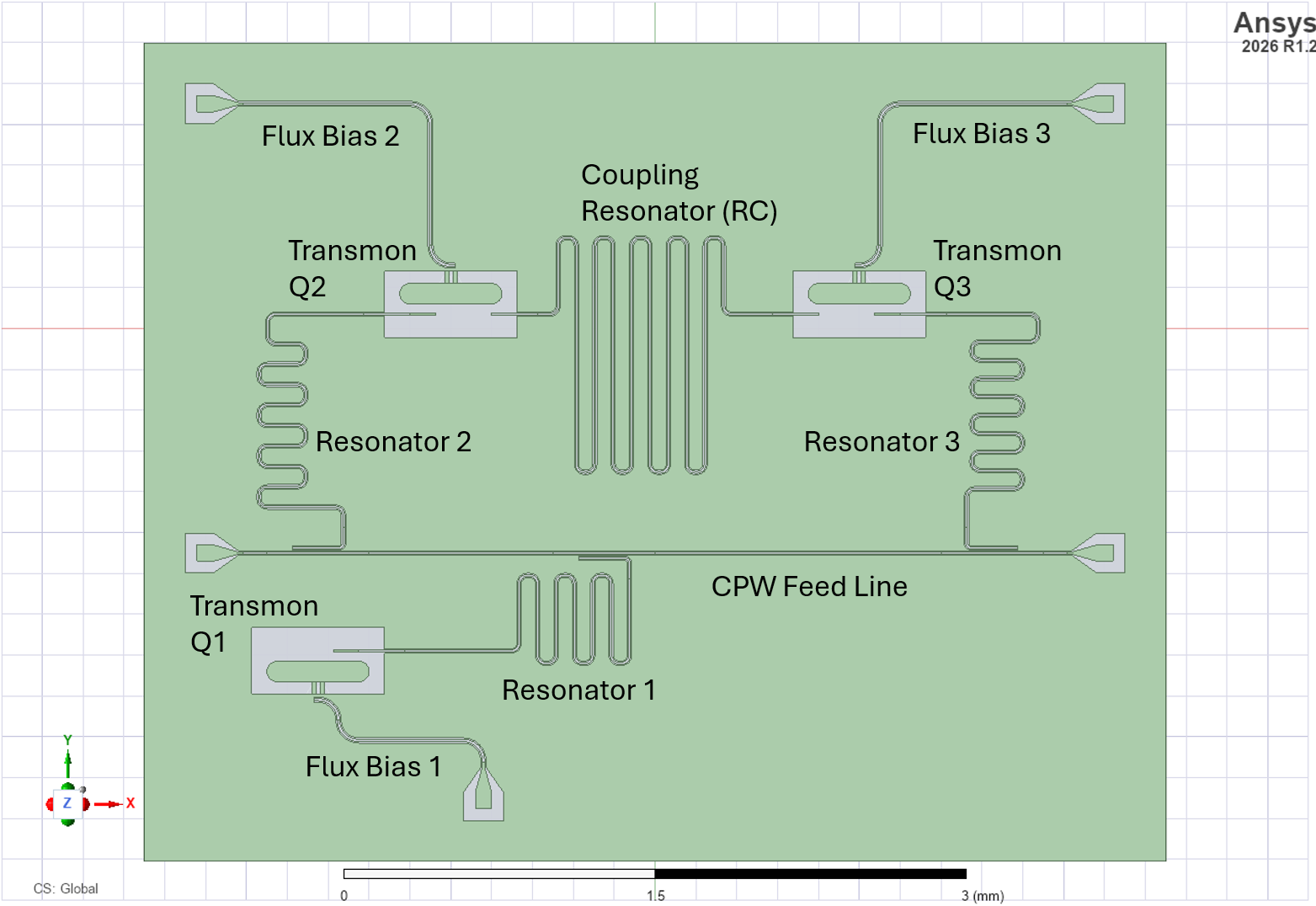}
\caption{The top view of a superconducting qubit chip rendered in HFSS, with the components identified. This is a 3-qubit chip to be designed in this paper.}
\label{fig:1_full_chip}
\end{figure}

Superconducting (SC) qubit technology is one of the most promising technologies to enable large-scale integration because it can reuse the silicon fabrication \cite{vandamme2024}\cite{montanezbarrera2025}\cite{google2025}\cite{Wong_review}. A superconducting qubit quantum chip contains qubits and various components and usually operates in the GHz regime \cite{Wong_review}. It is nothing but a microwave circuit operating in the quantum regime at cryogenic temperatures (e.g., 10 $mK$), exploiting the quantization of electromagnetic waves (Fig.~\ref{fig:1_full_chip}). Moreover, it is shown that the quantum parameters can be deduced from classical microwave circuit parameters \cite{minev2021}. Therefore, the design of an SC qubit chip can be largely reduced to the design of a classical microwave chip. 

Various types of superconducting qubits have been proposed and studied in the literature. A summary of various types of superconducting qubits can be found in, e.g.,  \cite{Wong_review}, \cite{gyenis2021}, and \cite{devoret2013}; among them, the transmon qubit is one of the most important ones due to its strongly reduced sensitivity to charge noise and long coherence time \cite{koch2007}. Therefore, transmon qubits will be used as examples in this paper. 

In this paper, using the 3-qubit chip in Fig.~\ref{fig:1_full_chip} as an example, we provide step-by-step guidance and examples of basic qubit chip design from theory to simulation. The following will be discussed:

\begin{itemize}
  \item Design of a coplanar waveguide (CPW) for a feedline (Section~\ref{sec:TL}).
  \item Design of a quarter-wavelength resonator and its coupling to a feedline (Section~\ref{subsec: resonator analytical}).
  \item Design of transmon qubits (Section~\ref{sec:transmon_qubit}).
    \item Design of flux-tunable qubits and flux bias lines (Section~\ref{sec:fluxtunable}).
  \item Design of a qubit-resonator coupling (Section~\ref{sec:qubitreadout}).
    \item Design of a 3-qubit chip with qubits coupled by a half-wavelength resonator (Section~\ref{sec:3qubit}).  
\end{itemize}

In each Section, basic physics and microwave engineering theory will be discussed. Then numerical calculations and/ or finite-element simulations are performed. In this paper, Quantum Metal \cite{qiskitMetal} is used to create quantum chip layouts and perform energy participation ratio (EPR) calculations. Ansys High Frequency Structure Simulator (HFSS) \cite{ansysHFSS} is used for finite element simulations of eigenmodes and scattering matrices. Ansys Q3D 
Extractor \cite{ansysQ3D}, and Maxwell 3D Design \cite{ansysMaxwell} are used for capacitance and magnetostatic simulations, respectively. YouTube videos are produced to demonstrate the setup and simulation procedures for each section, and the Quantum Metal source code is available on GitHub. The playlist of the videos can be found in \cite{YouTube_playlist} and the source codes and HFSS projects can be found in \cite{GitHub_repo}.

\subsection{Overview of Superconducting Qubit Chip and its Design Flow}

Fig.~\ref{fig:1_full_chip} shows the superconducting qubit chip to be designed in this paper and the typical components in a superconducting quantum chip. Naturally, it has superconducting qubits (Q1, Q2, Q3), which are transmons in the example. A transmon has a large shunt capacitor and a Josephson junction (JJ). Each transmon is coupled to a resonator (R1, R2, R3). The resonators are coupled to a CPW feedline through which the qubit state can be determined. In this design, the feedline-resonator system can also be used to manipulate the qubits to implement qubit gates. A coupling resonator (RC) is also used to couple the top two qubits, enabling 2-qubit gates between Q2 and Q3. The qubits can be flux-tunable (with two JJs in a loop) so that their frequency can be changed by adjusting the flux. Flux bias lines are included to adjust the transmons frequencies. Magnetic flux is created when a current flows through. Finally, all readout and qubit manipulation signals are delivered into the left launch pad of the CPW feedline. For readout signals, they are detected in the right launch pad. All launch pads will be wire-bonded to a printed circuit board (PCB) for the chip to interact with other components and equipment.

The goals in an SC quantum chip design are to ensure that DiVincenzo’s criteria \cite{divincenzo2000} are not violated. That means qubits should have long coherence time, readout speed should be fast enough, and manipulation of qubits should be possible. The following is a suggested design procedure. Some steps have questions and answers based on this 3-qubit chip example. It will start with theoretical calculations followed by simulations. Iterations are usually required.

\begin{enumerate}
  \item Choose the architecture
  \begin{enumerate}
    \item How many qubits? Three.
    \item Which qubits are coupled? Top two.
    \item Which substrate? Silicon.
    \item What type of superconducting metal? Aluminum.
    \item What type of qubits? Flux-tunable transmons.
    \item How to perform readout? Dispersive readout.
    \item How to perform manipulation? Share the same feedline as the readout.
  \end{enumerate}
  \item What are the qubit and resonator frequencies? See Table~\ref{tab:3-qubit_design_eigenmode_sim}.
  \item Perform theoretical calculations
    \begin{enumerate}
    \item Design CPW to match the system impedance. The same design can be used for resonators and flux-bias lines.
    \item Design the length of the resonators based on the desired frequency and its type (e.g., quarter- or half-wave length).
    \item Design the shunt capacitance and Josephson inductance for the desired qubit frequency and anharmonicity.
    \item Design the flux-bias line and JJ loop to ensure possible frequency tuning in the desired range.
  \end{enumerate}
 \item Chip layout
 \item Perform Simulations
    \begin{enumerate}
    \item Perform scattering matrix simulations to verify the design of the feedline and its coupling to the resonator with a reasonable quality factor.
    \item Perform eigenmode simulations to ensure the resonator and qubit frequencies are close to the desired values. The quality factor should be checked to ensure the qubits have long coherence time.
    \item Perform EPR calculations to ensure cross-Kerr is sufficient for distinguishing $\ket{0}$ and $\ket{1}$ states.
    \item Perform magnetostatic simulation to ensure the flux through the JJ loop is large enough under a bias current well below the critical current density level.
  \end{enumerate}
\end{enumerate}

A successful design likely will involve simulations in the early phase. For example, shunt capacitance extraction should be performed as early as layout design.

\section{Design of Coplanar Waveguide (CPW) Feedline}\label{sec:TL}

\subsection{Theory}
\subsubsection{Transmission Line}
At low frequency, an electro-magnetic (EM) wave has a wavelength, $\lambda$, much larger than the size of electrical components. Therefore, the electrical components can be treated as lumped elements because the EM wave properties (e.g., electric field magnitude and phase) vary slowly (almost constant) over the components. However, at higher frequencies ($f$), the wavelength becomes comparable to electrical components and Maxwell's equations need to be solved. For example, at a frequency of 5 GHz, the corresponding wavelength in vacuum is about 60 mm. In a non-vacuum environment, with an effective dielectric constant, $\epsilon_\mathrm{eff}$, the wavelength is given by,
\begin{equation}
    \lambda = \frac{c}{f\sqrt{\epsilon_\mathrm{eff}}}.
    \label{eq: wavelengtheff}
\end{equation}
where $c$ is the speed of light in the vacuum. For $\epsilon_\mathrm{eff}=6.45$, the wavelength will be reduced to 23.6 $mm$, which is comparable to the scale of the chip and its components shown in Fig.~\ref{fig:1_full_chip}. 

\begin{figure}[H]
\centering
\includegraphics[width=1\textwidth]{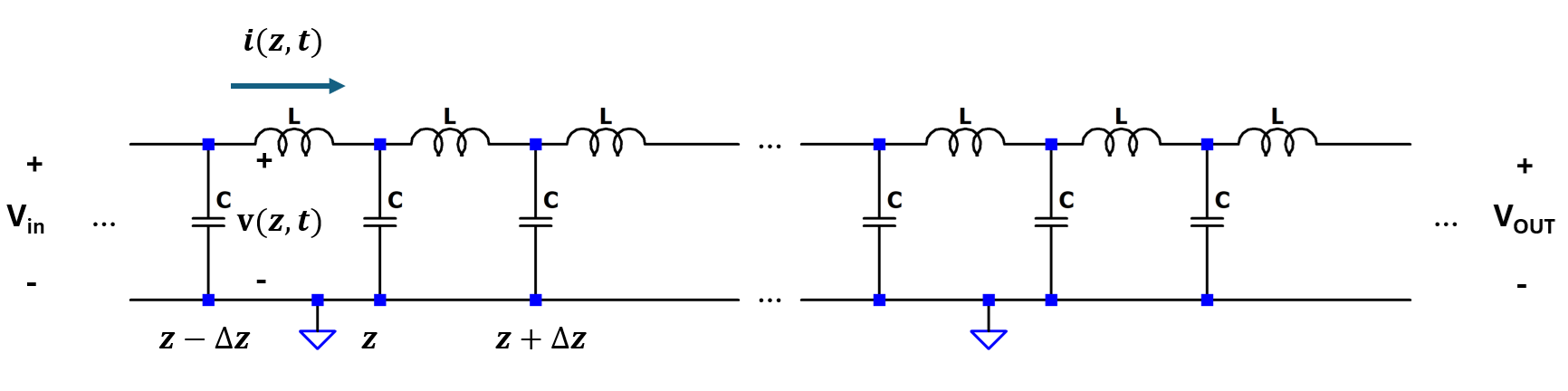}
\caption{Lumped-element transmission line model.}
\label{fig:LumpedTLModel}
\end{figure}

Transmission lines are components that need to be modeled "distributively". A transmission line has a conductive path for the signal and a return path connected to ground. To transmit an EM wave "smoothly" (i.e., without reflection), the transmission line needs to have an impedance matching the characteristic impedance, $Z_{0}$, of the system (including the source of the EM wave, e.g., 50 $\Omega$). For a transmission line, one may use a lumped-element model without solving Maxwell's equations explicitly. Fig.~\ref{fig:LumpedTLModel} shows the lumped-element transmission line model for a lossless transmission line. 


In the figure, $L$ is the inductance per unit length and $C$ is the capacitance per unit length, due to the coupling between the signal and ground metals.

The model in Fig.~\ref{fig:LumpedTLModel}  allows one to establish the relationship between the voltage and current at different locations, $z$, along the circuit. The locations between adjacent nodes differ by $\Delta z$. Taking the limit $\Delta z \xrightarrow{} 0$, we obtain the \textit{Telegrapher Equations} \cite{pozar2012}.
\begin{equation}
\begin{cases}
\frac{\partial v(z,t)}{\partial z} &= - L \frac{\partial i(z,t)}{\partial t} \\
\frac{\partial i(z,t)}{\partial z} &= -C \frac{\partial v(z,t)}{\partial t}
\end{cases},
\label{eq:TL eqns}
\end{equation}
where $v(z,t)$ and $i(z,t)$ are the voltage and current at location $z$ at time $t$, respectively.
For the sinusoidal steady-state condition, we can represent the current and voltage in the phasor domain with angular frequency, $\omega$, using:

\begin{equation}
\begin{cases}
v(z,t) &= \operatorname{Re}\{V(z) e^{j\omega t}\} \\
i(z,t) &= \operatorname{Re}\{I(z) e^{j\omega t}\}
\end{cases}.
\label{eq: phasors}
\end{equation}

Eq. (\ref{eq:TL eqns}) thus become two wave equations,

\begin{equation}
\begin{cases}
\frac{d^{2}V(z)}{dz^2} + \beta^{2} V(z) &= 0 \\
\frac{d^{2}I(z)}{dz^2} + \beta^{2} I(z) &=0 
\end{cases},
\label{eq: wave eqns}
\end{equation}
where $\beta=\omega\sqrt{LC}$. This is also the \textit{wavenumber} based on its role in the wave equations, and thus the wavelength is given by:

\begin{equation}
    \lambda = \frac{2\pi}{\beta}.
    \label{eq: wavelength}
\end{equation}

The solutions to the wave equations are,

\begin{equation}
\begin{cases}
V(z) &= V_{0}^{+} e^{-j\beta z} + V_{0}^{-} e^{j\beta z}\\
I(z) &= I_{0}^{+} e^{-j\beta z} + I_{0}^{-} e^{j\beta z} 
\end{cases},
\label{eq: wave eqn solns}
\end{equation}
where the total voltage (current) is the sum of the forward ($+$) and backward ($-$) voltage (current) waves. $V_{0}^{+}$, $V_{0}^{-}$, $I_{0}^{+}$, and $I_{0}^{-}$ are the complex amplitudes of forward voltage, backward voltage, forward current, and backward current waves at $z=0$, respectively.

The characteristic impedance of the transmission line, $Z_{0}$ is defined as 
\begin{equation}
    Z_{0}=\frac{V_{0}^{+}}{I_{0}^{+}} = -\frac{V_{0}^{-}}{I_{0}^{-}} .
    \label{eq: forward and backward waves}
\end{equation}
It can be shown that $Z_{0} = \sqrt{L/C}$ \cite{pozar2012}. 

In reality, a transmission line cannot be infinitely long and will be terminated by a load impedance, $Z_L$. The traveling wave will be reflected when it hits the load. If the load is at $z=0$, the ratio of the return wave complex amplitude to the forward one at the load is given by \cite{pozar2012}

\begin{equation}
\Gamma=\frac{V_{0}^{-}}{V_{0}^{+}} =\frac{Z_{L} -Z_{0}}{Z_{L} + Z_{0}}.
    \label{eq: gamma}
\end{equation}
It can be seen that if $Z_{L} =Z_{0}$ (i.e., with impedance matching), $\Gamma=0$ and there will be no reflection.

\begin{figure}[H]
\centering
\includegraphics[width=1\textwidth]{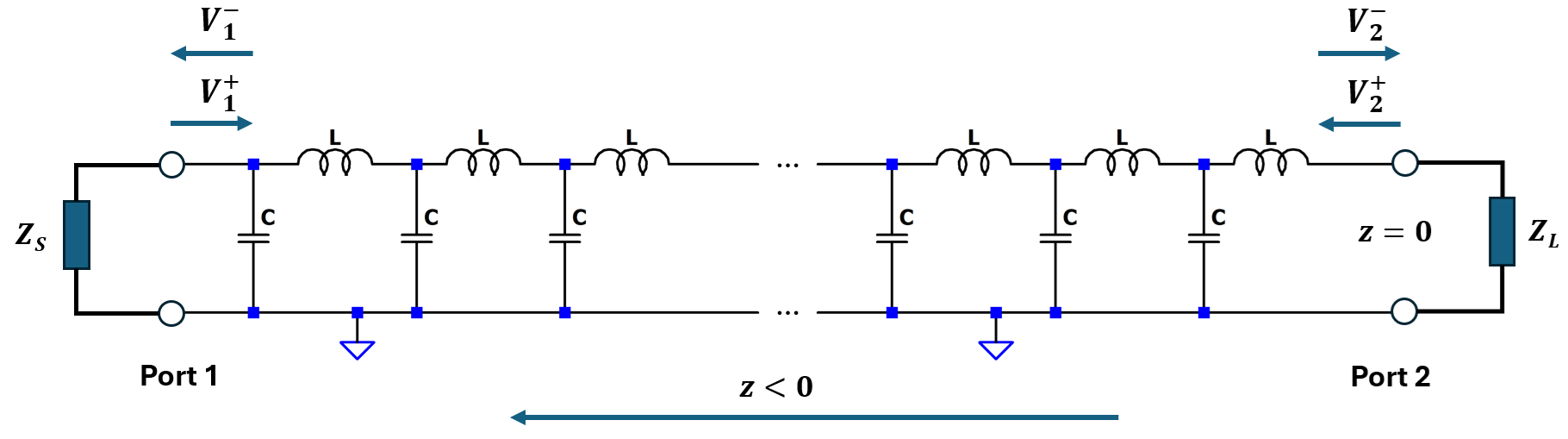}
\caption{Transmission line as a 2-port network connected to a source and a load impedance. Note that $z=0$ is at the load and $z$ is more negative in the left direction.}
\label{fig:LumpedTLModel2}
\end{figure}

Fig.~\ref{fig:LumpedTLModel2} shows a transmission line connected by a source impedance on the left ($Z_S$) and a load impedance on the right ($Z_L$). It can also be viewed as a 2-port network. The incident waves at the ports 1 and 2 are $V_{1}^{+}$ and $V_{2}^{+}$, respectively. The reflected waves are $V_{1}^{-}$ and $V_{2}^{-}$. They represent the complex amplitudes of the waves injected into the ports at the port locations and the complex amplitudes of the corresponding reflected waves. Therefore, $V_{2}^{+}=V_{0}^{-}$ and $V_{2}^{-}=V_{0}^{+}$. But it should be noted that Eq.~(\ref{eq: gamma}) is for calculating the reflected wave-to-incident wave amplitude ratio for a right-traveling incident wave (i.e., $V_{0}^{+}$). This is different from the definition of the incident wave in the two-port network (i.e., $V_{2}^{+}$).
One can correlate the four amplitudes through a scattering matrix with four scattering parameters through

\begin{equation}
\begin{bmatrix}
V_1^-\\[2pt]
V_2^-
\end{bmatrix}
=
\begin{bmatrix}
S_{11} & S_{12}\\[2pt]
S_{21} & S_{22}
\end{bmatrix}
\begin{bmatrix}
V_1^+\\[2pt]
V_2^+
\end{bmatrix}.
    \label{eq: s-matrixfull}
\end{equation}

The scattering parameters can be found by 

\begin{equation}
S_{11}
= \left.\frac{V_1^-}{V_1^+}\right|_{V_2^+=0},
\qquad
S_{12}
= \left.\frac{V_1^-}{V_2^+}\right|_{V_1^+=0},
\qquad
S_{21}
= \left.\frac{V_2^-}{V_1^+}\right|_{V_2^+=0},
\qquad
S_{22}
= \left.\frac{V_2^-}{V_2^+}\right|_{V_1^+=0}.
    \label{eq:smatrixelements}
\end{equation}

If $Z_0=Z_L=Z_S$, due to impedance matching, there will be no reflection at any point along the circuit for a wave traveling in any direction, as predicted by Eq.~(\ref{eq: gamma}) by setting $z=0$ at that point. Therefore, if a wave is injected into port 1 ($V^+_1$), one would expect $S_{11}=0$ due to no reflection ( $V^-_1=0$). All energy will propagate to and exist at port 2 without reflection at port 2 either ($V^+_2=0$). Therefore, $|S_{21}|=1$. It should be noted that in this case, only $|V^+_1|=|V^-_2|$ because there is a phase difference between them depending on the length $l$ of the transmission line (see Eq.~(\ref{eq: wave eqn solns})).

\subsubsection{Coplanar Waveguide Design}
\label{sec: CPW analytical}

\begin{figure}[H]
\centering
\includegraphics[width=1\textwidth]{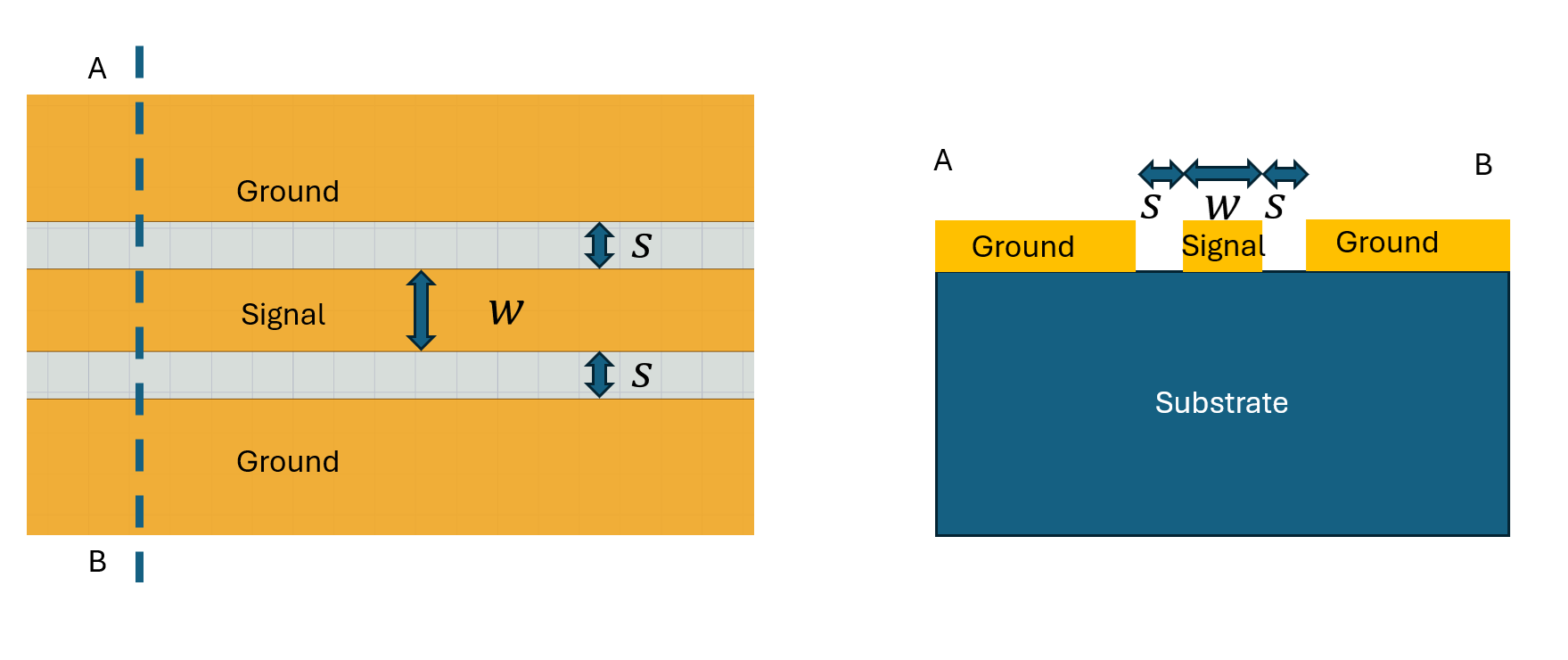}
\caption{Left: Top view of a portion of CPW. Right: Cross-section of a CPW at $A-B$.}
\label{fig:TLCrossSection}
\end{figure}

CPW is a convenient way to implement a transmission line on a quantum chip. A CPW is formed by a conducting line for carrying a signal with width $w$ shielded by two ground planes separated by a gap of $s$ from the signal line (Fig.~\ref{fig:TLCrossSection}).

The most important property of a CPW transmission line is its characteristic impedance, $Z_0$. It depends on $w$ and $s$. Its exact value can only be found by solving Maxwell's equations. However, with some approximations, it can also be found analytically. Firstly, we assume the metal has zero thickness (sheet). Secondly, the effective dielectric constant, $\epsilon_\mathrm{eff}$, can be approximated as the average of the dielectric constants above the CPW (i.e., air, $\epsilon_{air}$) and that below the CPW (i.e., silicon in this case, $\epsilon_{si}$). Therefore, 
\begin{align}
    \epsilon_\mathrm{eff} &= \frac{\epsilon_{si} + 
    \label{eq: dielectric constant1}
    \epsilon_{air}}{2}, \\
    &= \frac{11.9 + 1}{2} = 6.45.
    \label{eq: dielectric constant}
\end{align}

The dielectric constants in Eq.~(\ref{eq: dielectric constant}) are taken from the materials data table in HFSS.
According to \cite{goppl2008}, the characteristic impedance of the CPW can be found using:

\begin{equation}
    Z_{0} = \sqrt{\frac{\mu_{0}}{16 \epsilon_{0}\epsilon_\mathrm{eff}}} \frac{K(k_{0}')}{K(k_{0})}.
    \label{eq: Z0}
\end{equation}
where $k_{0} = \frac{w}{w + 2s}$ and $k_{0}' = \sqrt{1-k_{0}^{2}}$. $\mu_{0}$ and $\epsilon_{0}$ are the free space permeability and permittivity, respectively. $K(k)$ is the complete elliptic integral of the first kind defined by:

\begin{equation}
    K(k) = \int_{0}^{\pi/2} \frac{d\theta}{\sqrt{1-k^{2} \text{sin}^{2} \theta}}.
    \label{eq: K integral}
\end{equation}

One may also find it numerically using software or website such as the one in \cite{cpwCalc}. With $w$ = 10 \textmu m and $s$ = 5.806 \textmu m, by assuming the silicon substrate thickness to be $750$ \textmu m, $Z_{0}$ is found to be $49.568 \; \Omega$.

\subsection{Finite Element Simulation} \label{sec:FEM_TL}
\begin{figure}[H]
	\centering
	\includegraphics[scale=0.34]{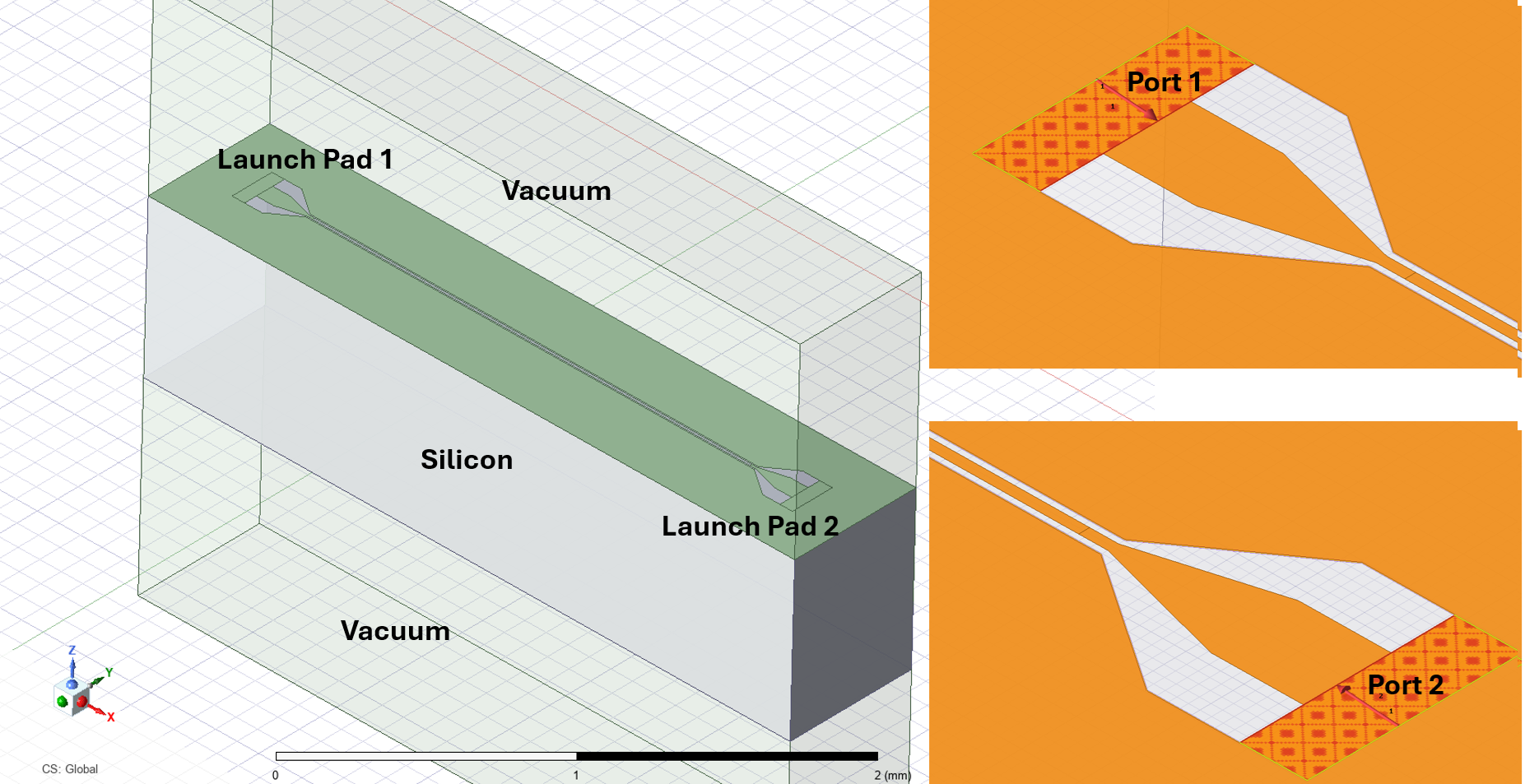}
	\caption{Left: CPW transmission line rendering in Ansys HFSS. Right: The launch pads are shown with the excitation ports and the corresponding integration lines highlighted.}
    \label{fig: transmission-line-rendering}
\end{figure}

Now, a finite element method (FEM) simulation tool will be used to analyze the CPW we have designed in \ref{sec: CPW analytical}. The goal is to make sure that $Z_{0}$ is about $50~\Omega$ for the chosen $w$ and $s$. Based on the discussion of Eq.~(\ref{eq:smatrixelements}), if the source and load impedance are the same as $Z_0$, $S_{11}=0$ and $|S_{21}|=1$. Therefore, we will sweep $Z_L$ and $Z_S$ together with $Z_S=Z_L$ to find out which value gives the minimum  $|S_{11}|$, and that $Z_L$ will also be $Z_0$. Solution type \textit{HFSS with Hybrid and Arrays} with \textit{Network Analysis} and \textit{Modal} options will be used by attaching lumped ports to both ends of a CPW transmission line.  Fig.~\ref{fig: transmission-line-rendering} shows the structure. The Jupyter notebook for creating the structure and the final structure for HFSS simulation can be found in \cite{Github_TL}. The corresponding YouTube video can be found in~\cite{YouTube_TL}.

In the final quantum chip, the transmission line in a quantum chip needs to be connected to the outside world. This is achieved by including launch pads at both ends, which will be connected to other electronics through wire bonding. Fig.~\ref{fig: transmission-line-rendering} shows the design of the launch pads. In HFSS simulation, a rectangle is created between the ground plane and the signal as a lumped port. To assign an object as a lumped port, right click the rectangle $>>>$ Assign Excitation $>>>$ Port $>>>$ Lumped Port. Then an impedance value representing the port impedance needs to be entered. The port impedance is equivalent to the effective impedance of the connected electronics seen by the transmission line and is the $Z_L$ (or $Z_S$) in Fig.~\ref{fig:LumpedTLModel2}. Next, an integration line is defined connecting the ground plane to the signal line  (Fig.~\ref{fig: transmission-line-rendering}). Since we plan to sweep the port impedance, in the lumped port setup, it is crucial to select \textit{Do Not Renormalize}.

Afterward, we need to create a \textit{solution setup}. This can be done by navigating to the Project Manager $>>>$ Right-click \textit{Analysis} $>>>$ Add Solution Setup $>>>$ Advanced. We select \textit{Broadband} and choose a frequency range of signals that we anticipate to expose to our quantum chip. In this case, we expect the chip to operate in the $4-8 \text{ GHz}$ range. The \textit{Maximum Number of Passes} is set to 45. Each pass is another opportunity for the simulation to do mesh refinement in an attempt to optimize for more accurate scattering parameters. Lastly, we configure the scattering convergence criteria, \textit{Maximum Delta S} which is the maximum change in our scattering parameters the code targets before terminating the simulation. If the target convergence criteria is not met, the simulation will pass through all \textit{Maximum Number of Passes}.

Navigating to the \textit{Adaptive Options} tab, we set the \textit{Maximum Refinement per Pass} to be $20\%$. This is decreased from the default $30\%$ to give the simulation a gentler, steadier convergence to our cost function if it isn't decreasing monotonically. We also define the \textit{Minimum Number of Passes} the simulation will perform before enabling the convergence termination criteria. Finally, \textit{Minimum Converged Passes} is the number of consecutive passes that the convergence criteria must remain satisfied before the simulation terminates. 

After \textit{Ok} is pressed, a frequency sweep will be prompted. Again, we want to configure the start and end frequencies to be within the operating frequency of our chip. We set the simulation type to \textit{Discrete}, the most accurate simulation type. Then, we can define how many points we want in our frequency sweep. Usually, a low number of discrete points are preferred as a discrete frequency sweep can be computationally expensive. All setup parameters described in this subsection are summarized in Table  \ref{tbl:drivenmodal}.

\begin{figure}[H]
	\centering
	\includegraphics[scale=0.33]{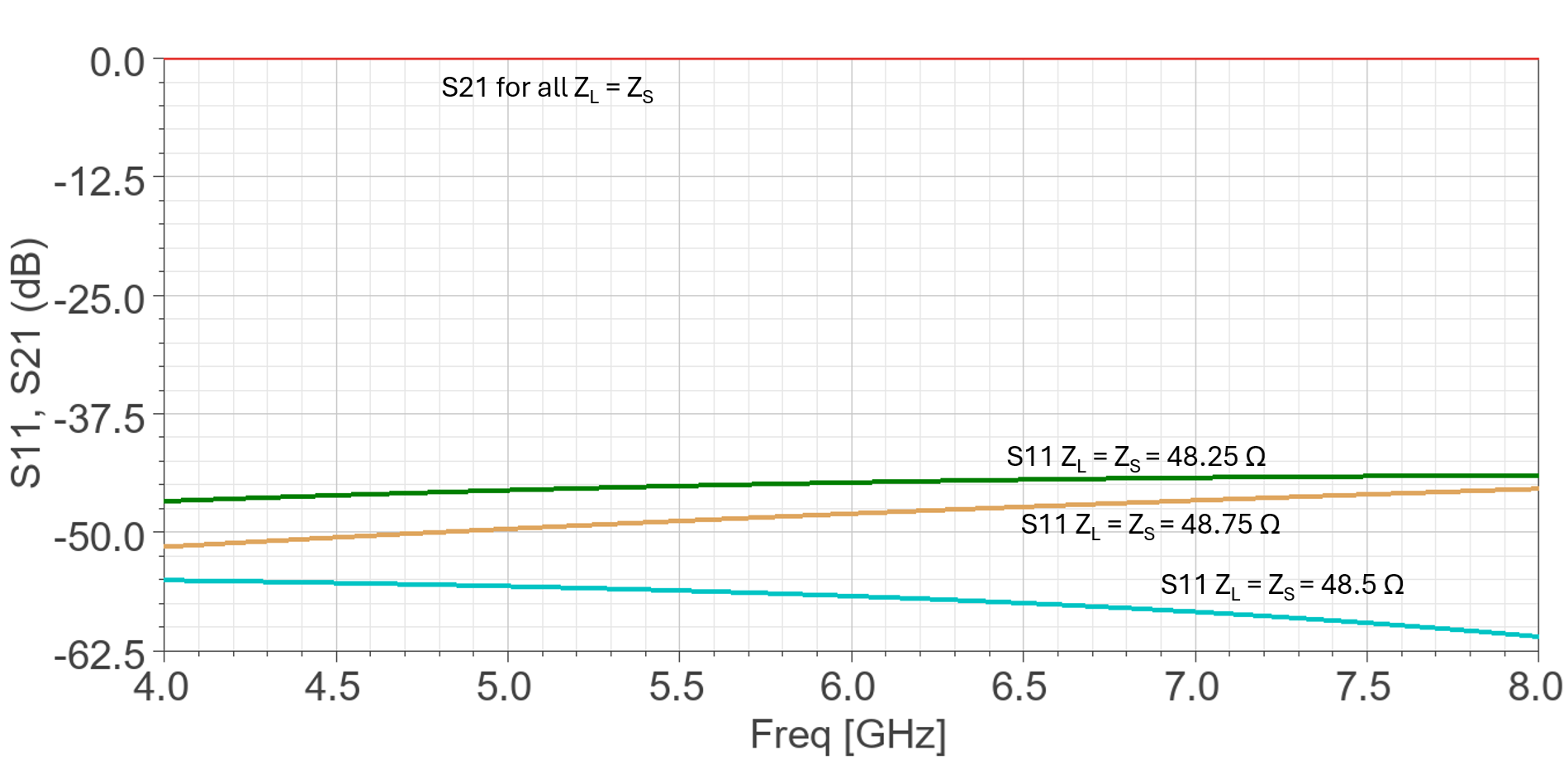}
	\caption{Scattering parameters, $S_{11}$ and $S_{21}$, as a function of frequency at different $Z_L=Z_S$ values.}
    \label{fig:TLHFSS}
\end{figure}

Fig.~\ref{fig:TLHFSS} shows $|S_{11}|$ and $|S_{21}|$ as a function of frequency at different $Z_L=Z_S$ values. They are expressed in dB which is $20\log_{10}|S_{ij}|$. $S_{21}$ is 0~dB in all ranges (equivalent to 1). That means almost all energy is transferred from Port 1 to Port 2. $S_{11}$ is essentially very small. The smallest one is when $Z_L=Z_S=48.5~\Omega$. Therefore, we can conclude that the CPW design has $Z_0=48.5~\Omega$.

\begin{table}[H]
\centering
\refstepcounter{table}
\addcontentsline{lot}{table}{Table \thetable\quad Transmission Line Impedance Matching Driven Modal Setup}
Table \thetable\\[0.5ex]
Transmission Line Impedance Matching Simulation Setup\\[1.5ex]
\label{tbl:drivenmodal}

\begin{tabular}{|l|l|}
\hline
\textbf{Parameter} & \textbf{Value} \\
\hline
Do Not Renormalize & True \\
\hline
Low Frequency (Start) & 4 GHz \\
\hline
High Frequency (End) & 8 GHz \\
\hline
Maximum Number of Passes & 45 \\
\hline
Maximum Delta S & 0.004 \\
\hline
Maximum Refinement per Pass & 20\% \\
\hline
Minimum Number of Passes & 35 \\
\hline
Minimum Converged Passes & 3 \\
\hline
Sweep Type & Discrete \\
\hline
Points & 21 \\
\hline
\end{tabular}
\end{table}

\subsection{Suggested Exercises}

\begin{itemize}
  \item Keep $w$ and $s$ ratio the same. Using the calculator in \cite{cpwCalc}, find whether $Z_0$ will change. Why?
  \item Design a transmission line with $Z_0=75~\Omega$. This is another typical system impedance used in TV, although not used for quantum chips.
\end{itemize}

\section{Design of CPW Quarter-Wavelength Resonator and its Coupling to a Feedline}\label{subsec: resonator analytical}

\subsection{Theory}
\subsubsection{Design of CPW Quarter-Wavelength Resonator}

The transmission line in Fig.~\ref{fig:LumpedTLModel2} can form a resonator if its $Z_S$ and $Z_L$ are terminated by open or short circuits. Both open and short circuits act as total reflectors. Based on Eq.~(\ref{eq: gamma}), when $Z_L \to \infty$, $V_0^-=V_0^+$. The wave is totally reflected ($\Gamma=1$) at the same phase as the outgoing wave. It thus has a constructive interference with a peak voltage at the open end (Eq.~(\ref{eq: wave eqn solns})) but with zero total current. Similarly, it is also totally reflected ($\Gamma=-1$) and has a destructive interference at a short end with zero voltage and maximum current. This can be seen by setting $z=0$ in Eq.~(\ref{eq: wave eqn solns}) and using the relationship in Eq.~(\ref{eq: forward and backward waves}).
\begin{figure}[H]
\centerline{\includegraphics[width=1\textwidth]{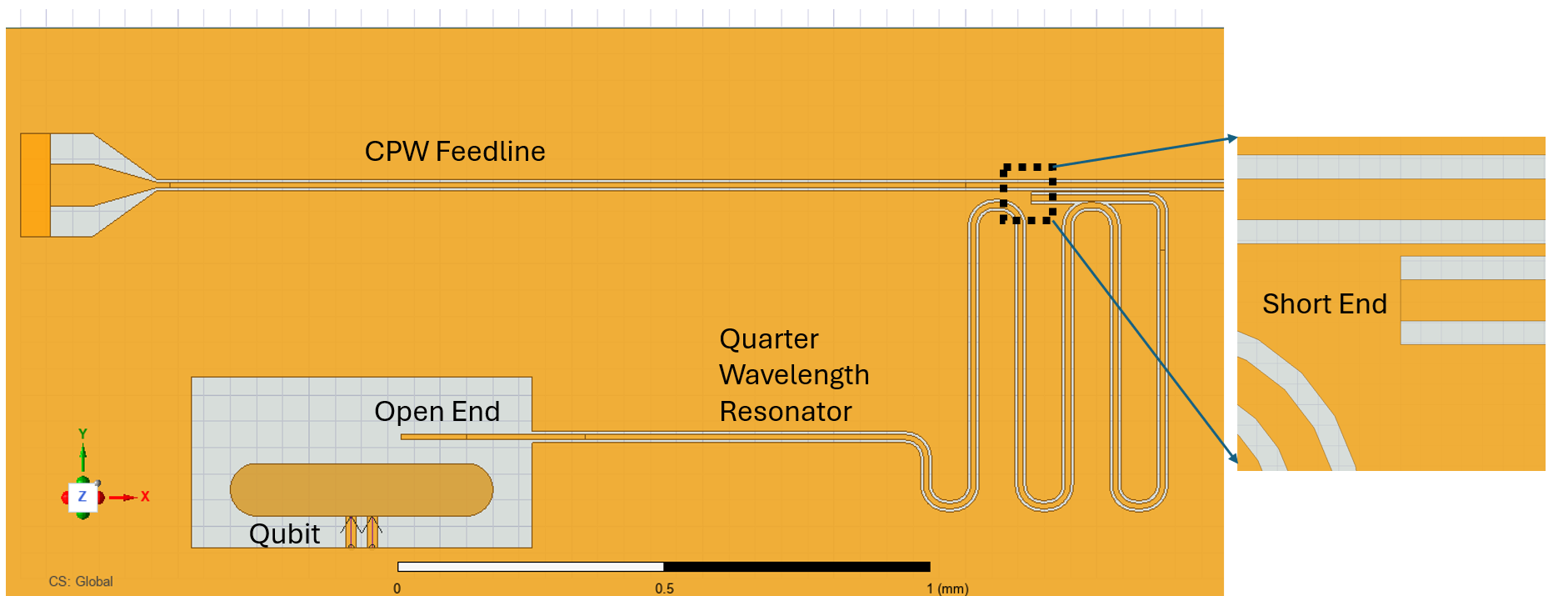}}
\caption{Quarter-wavelength ($\lambda/4$) resonator geometry. It is inductively coupled to the feedline at the short end and capacitively coupled to a qubit at the open end.}
\label{fig: quarter-wavelength-resonator-ansys-geometry}
\end{figure}
An EM wave of a certain frequency, the resonant frequency $f_r$ (or $\omega_r=2\pi f_r$ in radians), can be stored in the resonator. It should be noted that light is an EM wave; therefore, it is common to describe the amount of energy stored as the number of photons stored in the resonator, despite the fact that the frequency is in the GHz regime and they are not visible. The reason that it can store photons is that a standing wave will form. For example, if $Z_S=0~\Omega$ and $Z_L \to \infty$, a photon with a wavelength $ \lambda$ equal to four times the transmission line length $l$ will be stored. This is called a quarter-wavelength ($\lambda/4$) resonator (Fig.~\ref{fig: quarter-wavelength-resonator-ansys-geometry}).

\begin{figure}[H]
	\centering
	\includegraphics[scale=0.3]{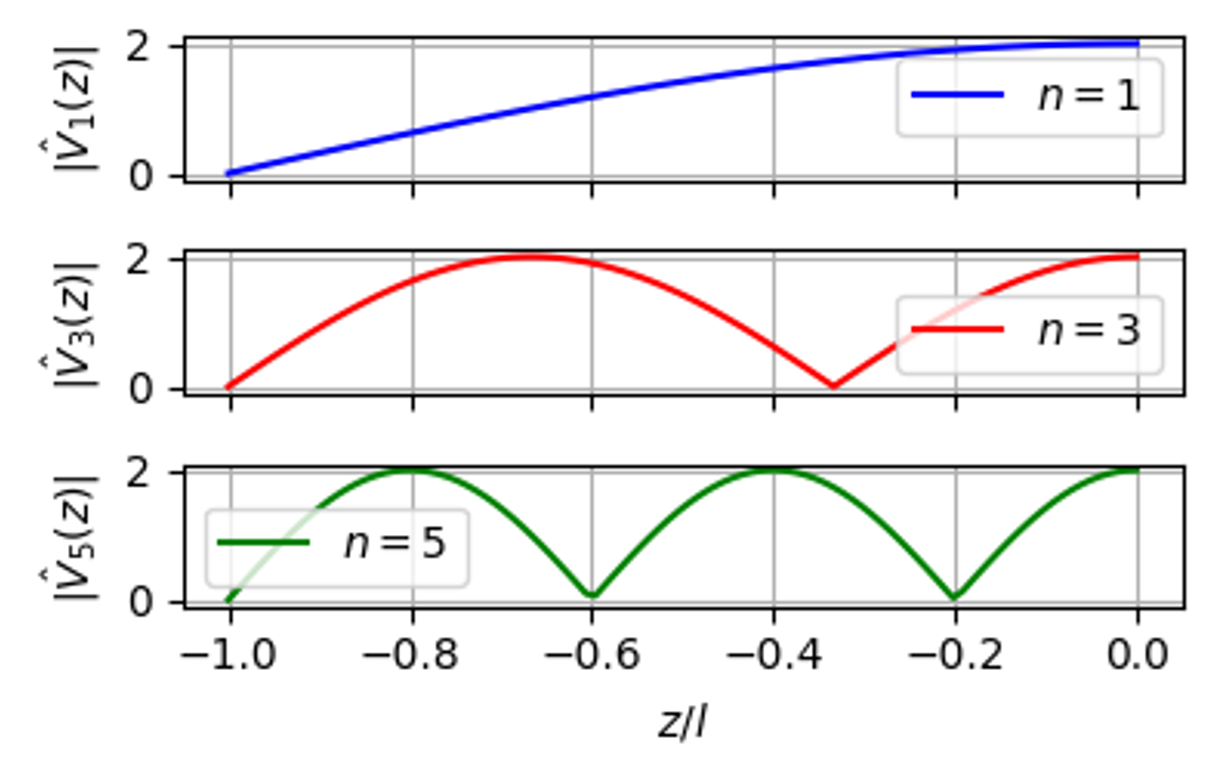}
	\caption{Voltage profiles of the first 3 modes in a quarter-wavelength resonator.}
    \label{fig: quarter-wavelength-analytical-plots}
\end{figure}

A resonator can store photons of different frequencies as it fulfills the boundary conditions that it has maximum voltage at the open end and zero voltage at the short end. Fig.~\ref{fig: quarter-wavelength-analytical-plots} shows the voltage profile of the lowest three modes in a quarter-wavelength resonator. The wavelengths are given by

\begin{equation}
    n\frac{\lambda}{4} = l ; \; n\in\text{odd integer},
    \label{eq: beta-l resonator}
\end{equation}
where $\lambda$ is the wavelength in the resonator. The corresponding frequencies are thus,

\begin{equation}
    f = \frac{c}{\sqrt{\epsilon_\mathrm{eff}}\lambda}
    =\frac{cn}{4l\sqrt{\epsilon_\mathrm{eff}}}
    ; \; n\in\text{odd integer}.
    \label{eq: quarter-wavelength resonator frequency spectrum}
\end{equation}

\subsubsection{Coupling of a $\lambda/4$ Resonator Coupled to a Feedline}

As shown in Fig.~\ref{fig: quarter-wavelength-resonator-ansys-geometry}, a resonator is coupled to a feedline to facilitate qubit measurement and manipulation. For a $\lambda/4$ resonator, it is coupled inductively to the feedline at the short side and capacitively to a qubit at the open end. The scattering parameters of the feedline will be affected by the resonator.

\begin{figure}[H]
	\centering
	\includegraphics[scale=0.3]{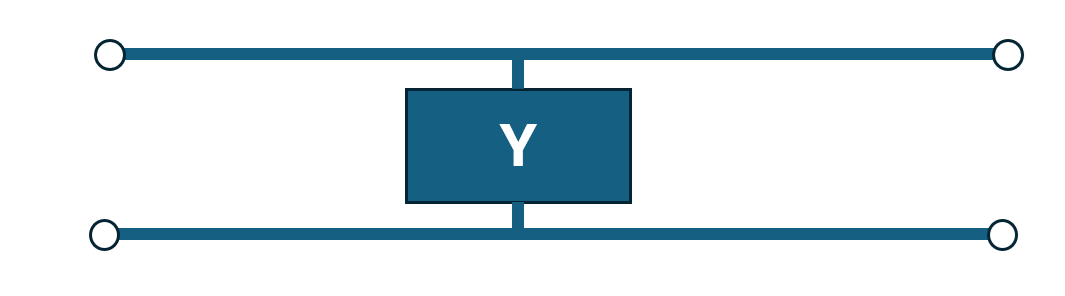}
	\caption{The equivalent network when the feedline is coupled to a resonator.}
    \label{fig:Y-network}
\end{figure}

The system can be modeled as in Fig.~\ref{fig:Y-network}. The admittance $Y$ is the effective impedance of the $\lambda/4$ resonator seen by the transmission line through inductive coupling.

The admittance of a $\lambda/4$ resonator can be derived using Eq.~(\ref{eq: wave eqn solns}) and Eq.~(\ref{eq: forward and backward waves}) by setting $z=0$ at the open end (this is equivalent to setting $Z_L\to \infty$ at the right side of Fig.~\ref{fig:LumpedTLModel2}). Then $\Gamma=1$ at $z=0$ gives $V_0^+=V_0^-$ and $I_0^+=-I_0^-$. Therefore, at the short end ($z=-l$ at the left side of Fig.~\ref{fig:LumpedTLModel2} with $Z_S=0$),
\begin{equation}
\begin{cases}
V(-l) &= V_{0}^{+} (e^{j\beta l} + e^{-j\beta l})\\
I(-l) &= I_{0}^{+} (e^{j\beta l} -  e^{-j\beta l}) 
\end{cases}.
\label{eq: wave eqn solns_quarter}
\end{equation}
The admittance of the $\lambda/4$ resonator seen at the short end is, thus, $Y' = I(-l)/V(-l) = j\; \text{tan}(\beta l) / Z_{0}$. But the resonator is inductively coupled to the transmission line. Therefore, the feedline does not see $Y'$ but with a scaling factor. For simplicity, we will model the effective admittance seen by the feedline using a frequency-independent scaling factor $s$ (not to be confused with the CPW gap). Therefore, $Y= 2js\; \text{tan}(\beta l) / Z_{0}$, where the factor of 2 is included for convenience.

Based on \cite{pozar2012_ch4}, the $S_{21}$ in Fig.~\ref{fig:Y-network} is thus,
\begin{align}
    S_{21} &= \frac{2}{A + B/Z_{0} + CZ_{0} + D}, \nonumber\\
    &= \frac{2}{1 + 0/Z_0 + YZ_{0} + 1}, \nonumber\\
    &= \frac{1}{1 + js\; \text{tan}(\beta l)},
    \label{eq: S21-resonatorY}
\end{align}
where $A$, $B$, $C$, and $D$ are the $ABCD$-matrix elements, and their values for this circuit are substituted in the second row.

The transmitted power ratio, $|S_{21}|^{2}$, is,
\begin{equation}
    |S_{21}|^{2} = \frac{1}{1 + s^2\tan^{2} (\beta l)}.
    \label{eq : res-S21}
\end{equation}

When Eq.~(\ref{eq: beta-l resonator}) is satisfied, $\beta l = n\pi/2$ with $n$ being odd, we obtain $|S_{21}|^{2}=0$. That means when the frequency is at the resonator's resonant frequency, a right-traveling wave cannot be transmitted to the right port. This can be used to detect the frequency of the resonator. Assume we want to design a resonator with $f_r=6.86$ GHz, we set $l=4.42$ mm based on Eq.~(\ref{eq: quarter-wavelength resonator frequency spectrum}) and $\epsilon_\mathrm{eff}=6.1$. Fig.~\ref{fig:S21_Y_network} shows $|S_{21}|^2$ with various feedline-resonator coupling strengths at different frequencies.

\subsubsection{Quality Factor} \label{Quality_factor_section}
The quality factor, $Q_r$, of a resonator tells us how long it can hold the photons it stores, $T_r$. It is related to the number of oscillations, $N_r$, before the stored energy is dissipated. A resonator can lose its photons through parasitic resistance, coupling to dielectrics (defects) in the environment, and coupling to other components such as the feedline near it. In a high-quality superconducting qubit chip operating in the $mK$ regime, we assume the loss is primarily due to coupling to the feedline. $Q_r$ is defined as 
\begin{align}
    Q_r &= 2\pi f \frac{\text{(average energy stored)}}{\text{(energy loss / second)}},\nonumber\\
    &=2\pi f T_r,\nonumber \\
    &=2\pi N_r.
    \label{eq:Q-definition}
\end{align}
Therefore, the higher $Q_r$ is, the longer it takes, or the more oscillation cycles are required, for the system to lose its stored energy.

\begin{figure}[H]
	\centering
	\includegraphics[scale=0.3]{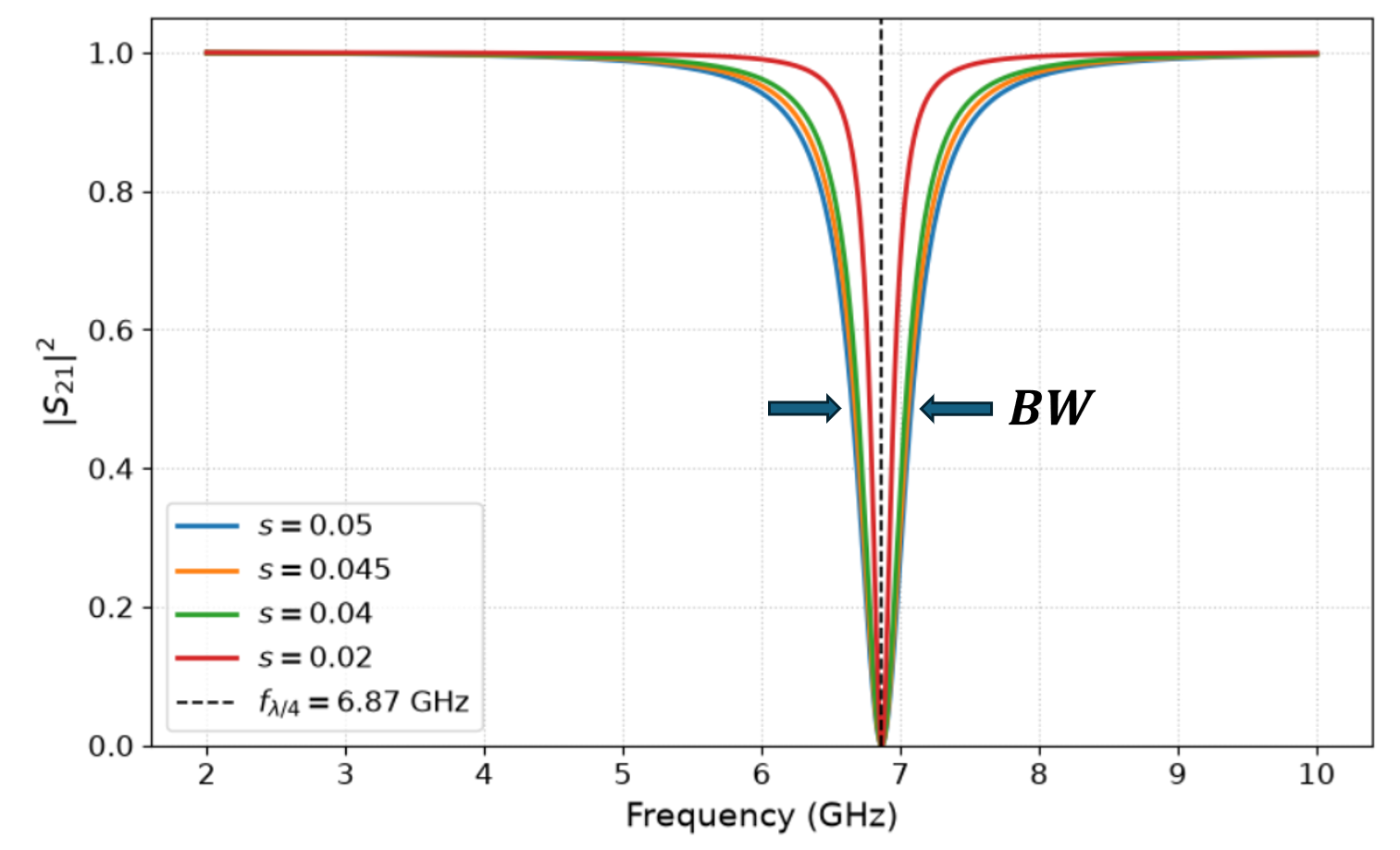}
	\caption{$|S_{21}|^{2}$ as a function of frequency and coupling strength, $s$, when a quarter-wavelength resonator is inductively coupled to the feedline using Eq.~(\ref{eq : res-S21}). $BW$ is defined using  $|S_{21}|^{2}=0.5$.}
\label{fig:S21_Y_network}
\end{figure}

$Q_r$ is also related to the bandwidth, $BW$, of the resonator. Fig.~\ref{fig:S21_Y_network} shows that when the coupling increases (easier to lose photons through the feedline), $BW$ is larger. Here, $BW$ is defined when $|S_{21}|^{2}=0.5$. It is related to $Q_r$ through this equation, 

\begin{align}
    BW &\approx \frac{f_r}{Q_r}.
    \label{eq:Q-definition_BW}
\end{align}

It should be noted that there are different $BW$ definitions for "notch-type" curves, like the one in Fig.~\ref{fig:S21_Y_network}. For a "peak-type", it is usually defined at the y-value when it is -3 dB from the peak.

If a resonator has other loss mechanisms, the total $Q_{total}$ can be found by,

\begin{equation}
    \frac{1}{Q_{total}} =\frac{1}{Q_{r}}+\frac{1}{Q_{other}}.
    \label{eq:Q-total}
\end{equation}

\subsection{Simulations}

To verify the design of the resonator and study its coupling to the feedline, HFSS simulations are performed. There are two types of simulations. The first one is the eigenmode simulation to find the resonant frequency and energy distribution. The other is scattering matrix simulation. The Jupyter notebook for creating the structure and the final structure for HFSS simulation can be found in \cite{Github_coupled}. The corresponding YouTube video can be found in~\cite{YouTube_coupled}.

\subsubsection{Eigenmode Simulation}
\label{sec: tl-res eigenmode}
In this section, we perform an eigenmode simulation to determine the fundamental frequency of the resonator and its quality factor. Assume the target frequency is $6.86 \; \text{GHz}$. Based on Eq.~(\ref{eq: quarter-wavelength resonator frequency spectrum}) with $n=1$ and by using $\epsilon_\mathrm{eff}=6.1$, we obtain $l=4.42$ mm. We used $\epsilon_\mathrm{eff}=6.1$ instead of 6.45 (see Eq.~(\ref{eq: dielectric constant})) to demonstrate that there are different methods to approximate the effective dielectric constant. This value is found by finding the ratio of the effective capacitance of a capacitive structure when the environment is all silicon versus half silicon \cite{YouTube_coupled}. This will not be discussed in detail here. The results are expected to be similar. The final layout is shown in Fig.~\ref{fig: quarter-wavelength-resonator-ansys-geometry}.

In order to simulate $Q$, port boundary conditions are applied to the ends of the transmission line with a lumped resistance of 50 $\Omega$. The construction is similar to what was done in Section~\ref{sec:FEM_TL} for scattering matrix simulations. However, for eigenmode simulations, there is a subtle difference in the setup: Right click the rectangle $>>>$ select \textit{Assign Boundary} $>>>$ Select \textit{Lumped RLC}.

To start an eigenmode simulation, select \textit{HFSS} $>>>$ \textit{Solution Type} and then the \textit{Eigenmode} option. Next, go to the \textit{Project Manager} $>>>$ Right click \textit{Analysis} $>>>$ \textit{Add Solution Setup}.

Then we set \textit{Number of Modes} to 6 and the eigenmode simulation will search for eigenfrequencies 
 greater than the \textit{Minimum frequency}, and will only record the first $6$. The \textit{Maximum Delta Frequency per Pass} is the convergence criterion that terminates the simulation if the difference in solution from pass to pass is within the specified difference. Select the \textit{Converge on Real Frequency} box to enable easier convergence. Under the adaptive options, we can specify our \textit{Maximum Refinement per Pass}, \textit{Maximum Number of Passes}, and \textit{Minimum Converged Passes}. All eigenmode simulation setup settings for this experiment are summarized by Table  \ref{tbl:eigenmodesetup}.

To view the simulation result containing the frequency and quality factor for each eigenmode, we can right-click \textit{Results} $>>>$ Select \textit{Solution Data} $>>>$ Select the \textit{Eigenmode Data} tab. The eigenmode data table will appear. Table \ref{tbl:eigenmodes} lists the eigenfrequencies and quality factors of each resonant mode.

Mode 1 is the fundamental frequency at $6.81$ GHz and is similar to the target of $6.86$ GHz. Mode 5, $20.79$ GHz, is the third harmonic with $n=3$ in Eq.~(\ref{eq: quarter-wavelength resonator frequency spectrum}), which is about 3 times the frequency of Mode 1 ($20.43$ GHz). The system and equipment will be limited to operating at a frequency less than $10$ GHz. Therefore, modes other than the $6.81$ GHz will be suppressed in the experiment.

Mode 1 also has $Q=7805.6$. Based on Eq.~(\ref{eq:Q-definition}), this means it takes the resonator $7805.6/(2\pi\times 6.81 \text{GHz}) = 182~$ns to lose a photon if one photon is stored. This is necessary to enable reasonable gate manipulation and readout time at the order of hundreds of ns. Based on Eq.~(\ref{eq:Q-definition_BW}), the $Q$ also corresponds to a $BW=\frac{6.81\times 10^9~\text{Hz}}{7805.6} = 0.87~\text{MHz}$

\begin{table}[H]
\centering
\refstepcounter{table}
\addcontentsline{lot}{table}{Table \thetable\quad Transmission Line–Resonator Eigenmode Setup}
Table \thetable\\[0.5ex]
Transmission Line–Resonator Eigenmode Simulation Setup\\[1.5ex]
\label{tbl:eigenmodesetup}

\renewcommand{\arraystretch}{1.3}

\begin{tabular}{|l|l|}
\hline
\textbf{Parameter} & \textbf{Value} \\
\hline
Minimum Frequency & 1 GHz \\
\hline
Number of Modes & 6 \\
\hline
Number of Passes & 18 \\
\hline
Maximum Delta Frequency per Pass & 0.5\% \\
\hline
Converge on Real Frequency & True \\
\hline
Maximum Refinement per Pass & 30\% \\
\hline
Minimum Number of Passes & 12 \\
\hline
Minimum Converged Passes & 3 \\
\hline
\end{tabular}
\end{table}

\begin{table}[H]
\centering
\refstepcounter{table}
\addcontentsline{lot}{table}{Table \thetable\quad Transmission Line–Resonator Eigenmode Results}
Table \thetable\\[0.5ex]
Transmission Line–Resonator Eigenmode Simulation Results\\[1.5ex]
\label{tbl:eigenmodes}

\renewcommand{\arraystretch}{1.3}

\begin{tabular}{|c|c|c|}
\hline
\textbf{Mode} & \textbf{Frequency (GHz)} & \textbf{Q factor} \\
\hline
       1  &  6.81&          7805.6\\
       \hline
       2  &  13.05 +      6.35 j   &       1.1\\
\hline
3   & 13.45&         551.0\\
\hline
       4 &   17.12    &       856.7\\

       \hline
       5  &  20.79&           3260.2\\
       \hline
       6    & 24.77&           560.6\\

\hline
\end{tabular}
\end{table}

\begin{figure}[H]
	\centering
	\includegraphics[scale=0.4]{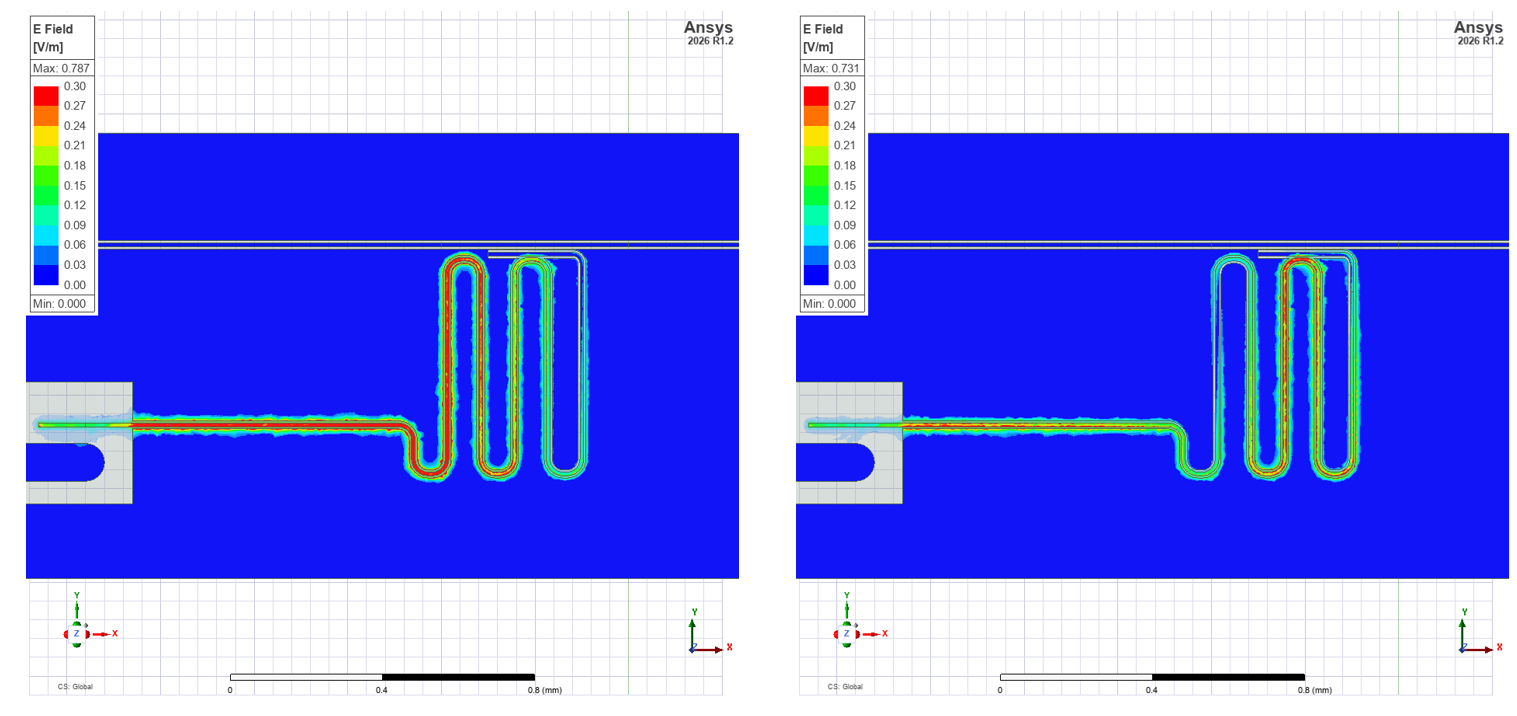}
	\caption{Left: Electric field distribution of Mode 1. Right: Electric field distribution of Mode 5.}
\label{fig:Eigenmode_Field}
\end{figure}

In Fig.~\ref{fig:Eigenmode_Field}, we plot the electric field distribution for Mode 1 and Mode 5. This is done by selecting the components and right-clicking on them, selecting \textit{plot fields}, then choosing \textit{E} for the electric field. It can be seen that Mode 1 has its peak electric field at the open end and minimum at the short end. Mode 5 has two peaks, with one at the open end and one closer to the short end. It also has a minimum at the short end. This agrees with the plot in Fig.~\ref{fig: quarter-wavelength-analytical-plots} for $n=1$ and $n=3$, respectively.

\begin{figure}[H]
	\centering
	\includegraphics[scale=0.35]{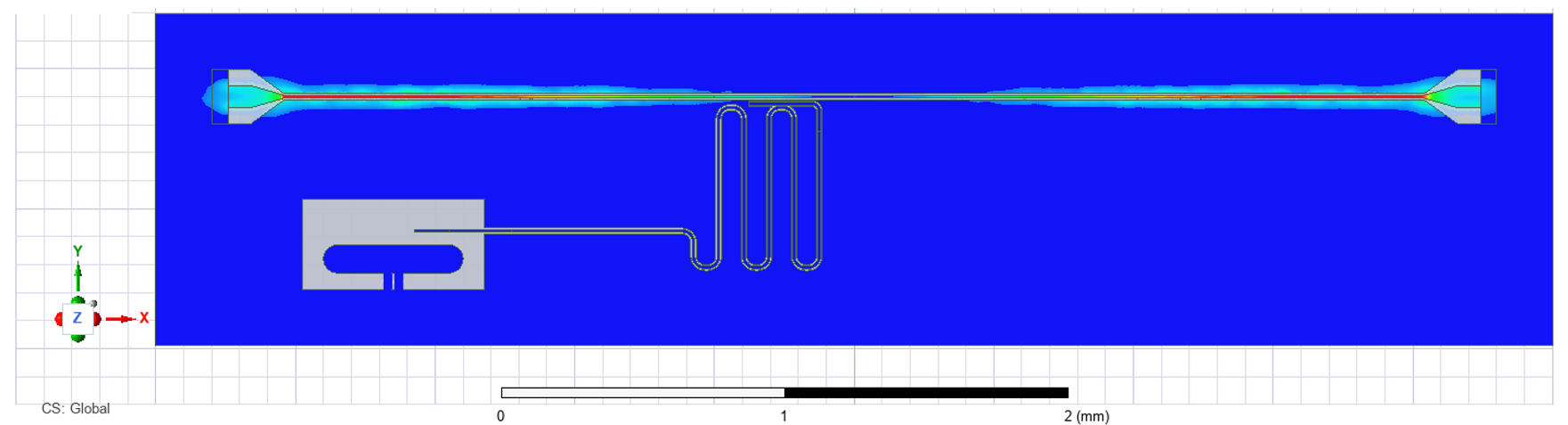}
	\caption{Electric field distribution of Mode 2.}
\label{fig:Eigenmode_feedline}
\end{figure}

It is also instructive to understand Mode 2, which has a large imaginary part and a small Q. This is a spurious mode caught in the simulation for the finite length of the feedline terminated by two launch pads. If the impedance matching is perfect, this mode will not exist. However, since it is not perfect, this mode is caught but with a very small $Q$. Based on Eq.~(\ref{eq:Q-definition}), if it were to store any photon, it takes only $\sim10$ ps to lose it. The frequency is similar to what is expected for a half-wavelength resonator with the feedline length of $4.04$ mm (excluding the launch pads). The half-wavelength resonator will be discussed in Section~\ref{sec:haflwavelength}. Fig.~\ref{fig:Eigenmode_feedline} shows the electric field distribution of Mode 2. The peak electric field can be seen at both ends, as it serves as a half-wavelength resonator with two open ends, but with very low $Q$ and high loss.

\subsubsection{Simulation Quality Check}

FEM simulations converge based on error criteria. This means that it assumes the simulation is converged when the change in certain calculated values (i.e., the error) is less than a certain preset percentage from one pass to the next. Therefore, there is no guarantee that the simulation will converge to the correct result. But we can use a few methods to inspect the quality of the simulation:

\begin{itemize}
    \item If available, compare the simulation result to an analytical solution.
    \item Check the structure for a dense enough mesh, especially in regions where the fields vary strongly.
    \item Verify monotonic convergence as a function of pass. That means the error is expected to decrease monotonically instead of oscillating. Compare it to the desirable max delta freq. \% in the \textit{Results} tab.

\end{itemize}

In this example, the resonator frequency (Mode 1) obtained is 6.81 GHz, which is pretty close to the expectation of $6.86 \; \text{GHz}$. But it should be noted that the analytical solution is also just an approximation. Moreover, it converges based on the "max delta freq." setting. 

\begin{figure}[H]
	\centering
	\includegraphics[scale=0.45]{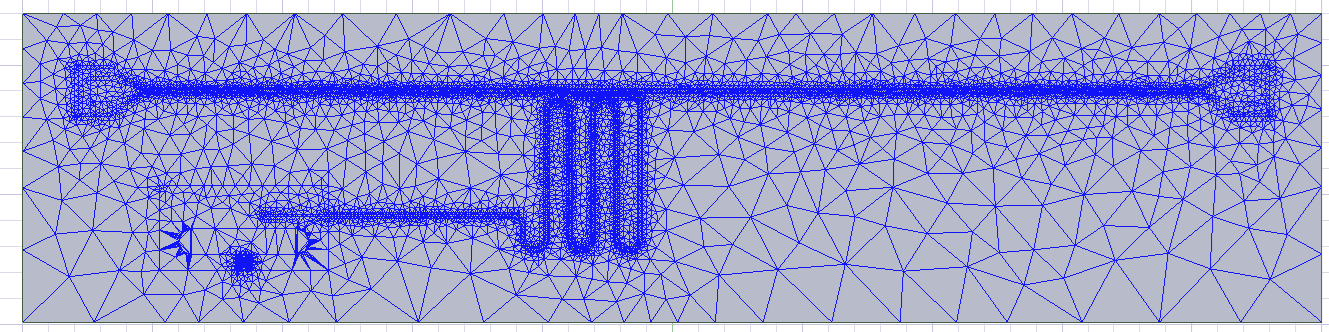}
	\caption{Mesh of a converged eigenmode simulation for the coupled feedline-resonator structure in Fig.~\ref{fig: quarter-wavelength-resonator-ansys-geometry}.}
    \label{fig: TL-res-mesh}
\end{figure}

In the event that we don't have an analytical model to rely on, we must rely on the other two techniques to check for simulation accuracy. Similar to the technique for plotting the fields on the chip in Section~\ref{sec: tl-res eigenmode}, we can select all materials (except vacuum) and then $>>>$ Right Click $>>>$ Select \textit{Plot Mesh} in order to assess the quality of our mesh (Fig.~\ref{fig: TL-res-mesh}). The simulation returns a dense mesh in the regions where we expect our signal to propagate and where the energy concentrates. In particular, the transmission line and the resonator have a sufficiently dense mesh.

\begin{figure}[H]
	\centering
	\includegraphics[scale=0.35]{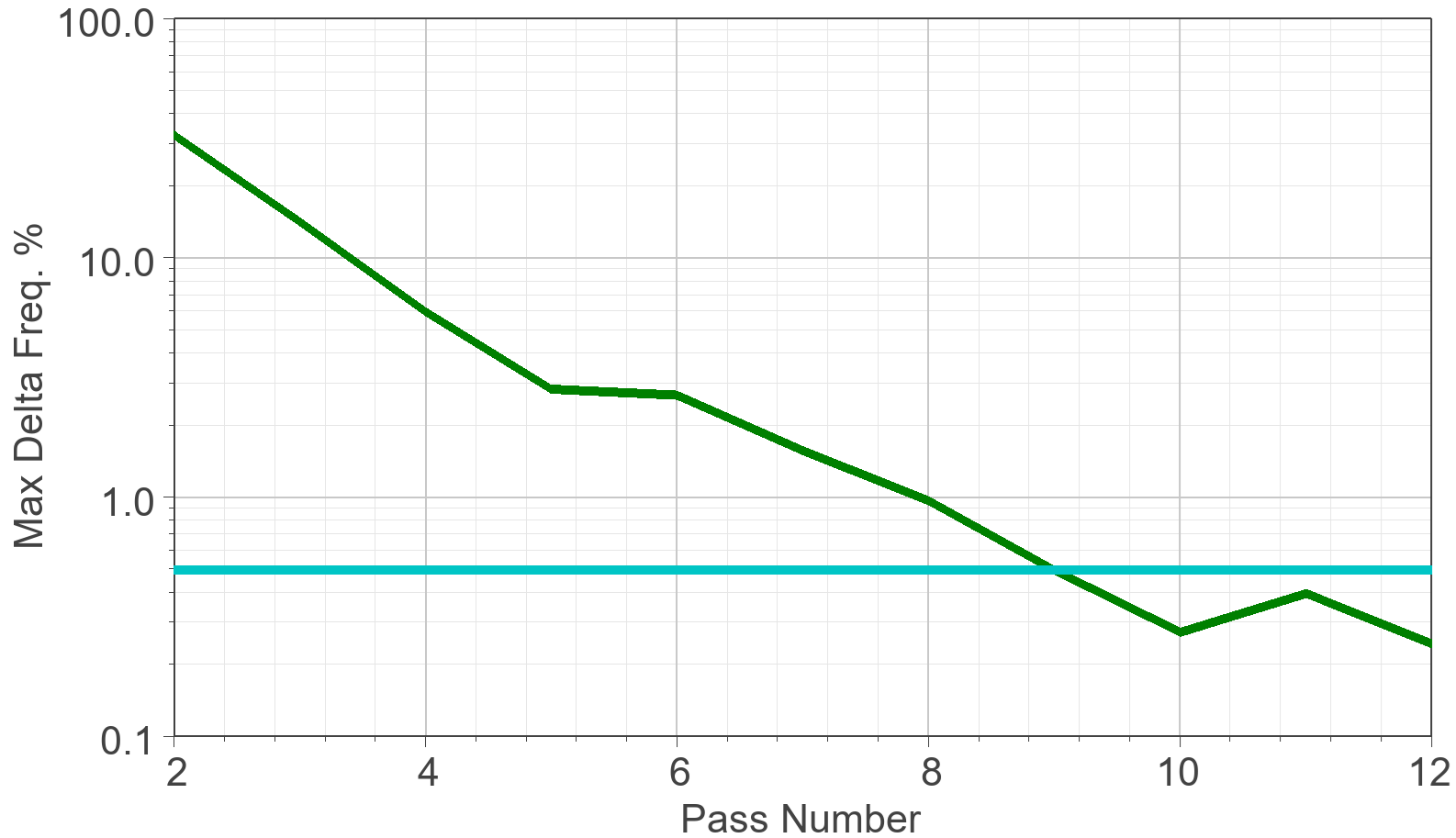}
	\caption{Coupled feedline-resonator eigenmode convergence information. The horizontal line shows the convergence criterion.}
    \label{fig: convergence-criteria-tl-resonator}
\end{figure}

To check for convergence behavior, we navigate to our \textit{Project Manager} $>>>$ Right click \textit{Results} $>>>$ Select \textit{Solution Data} $>>>$ Select the \textit{Convergence} $>>>$ Click the \textit{Plot} button. A plot of the Max Delta Freq \% vs pass will appear (Fig.~\ref{fig: convergence-criteria-tl-resonator}), along with other simulation metrics. It can be seen that the frequency difference (error) decreases almost monotonically. Also, when it has met the convergence criteria in Pass 10, it runs two more passes because "Minimum Converged Passes" is set to 3 in Table~\ref{tbl:eigenmodesetup}. 


\subsubsection{Scattering Matrix Simulation}
\label{sec: import mesh}

To perform scattering simulations of the coupled feedline-resonator system, we use the same methodology as in Section~\ref{sec:FEM_TL}.

\begin{table}[H]
\centering
\refstepcounter{table}
\addcontentsline{lot}{table}{Table \thetable\quad Transmission Line–Resonator Driven Modal Setup}
Table \thetable\\[0.5ex]
Coupled Feedline–resonator Scattering Matrix Simulation Setup\\[1.5ex]
\label{tbl:tlresonator_drivenmodal}

\renewcommand{\arraystretch}{1.3}

\begin{tabular}{|l|l|}
\hline
\textbf{Parameter} & \textbf{Value} \\
\hline
Coarse Frequency Range & 6.78 - 6.82 GHz \\
\hline
Fine Frequency Range & 6.8 - 6.803 GHz\\
\hline
Maximum Number of Passes & 3 \\
\hline
Maximum Delta S & 0.02 \\
\hline
Maximum Refinement per Pass & 30\% \\
\hline
Minimum Number of Passes & 3 \\
\hline
Minimum Converged Passes & 1 \\
\hline
Sweep Type & Interpolation \\
\hline
Points & 51 \\
\hline
Import Mesh & True \\
\hline
Source Design & Eigenmode Project \\
\hline
Simulate Source Design as Needed & False \\
\hline
Preserve Source Design Solution & True \\
\hline
\end{tabular}
\end{table}

We expect to observe a dip in the $|S_{21}|^2$ as shown in Fig.~\ref{fig:S21_Y_network}. However, this requires a very dense mesh. It is often difficult to rely on automatic meshing from pass to pass to obtain an appropriate mesh. Based on \cite{Dhillon_import}, one can import the eigenmode mesh as a starting point. We import the mesh from the eigenmode simulation in Fig.~\ref{fig: TL-res-mesh} as the starting mesh for the scattering simulation. To import the mesh from another project, we double-click the \textit{Solution Setup} $>>>$ navigate to the \textit{Advanced} tab $>>>$ \textit{Import Mesh}. The video in \cite{YouTube_Import} demonstrates the procedure. The scattering solution setup, including the mesh import, is summarized by Table  \ref{tbl:tlresonator_drivenmodal}.
We first select the frequency interval to be between $6.78 \text{ GHz}$ and $6.82 \text{ GHz}$, since we calculated from Section~\ref{sec: tl-res eigenmode} that the first resonator mode is $6.81 \text{ GHz}$, which lies within this domain. After that, we apply a second frequency range between $6.8 \text{ GHz}$ and $6.803 \text{ GHz}$ to further resolve the dip.

\begin{figure}[H]
	\centering
	\includegraphics[scale=0.3]{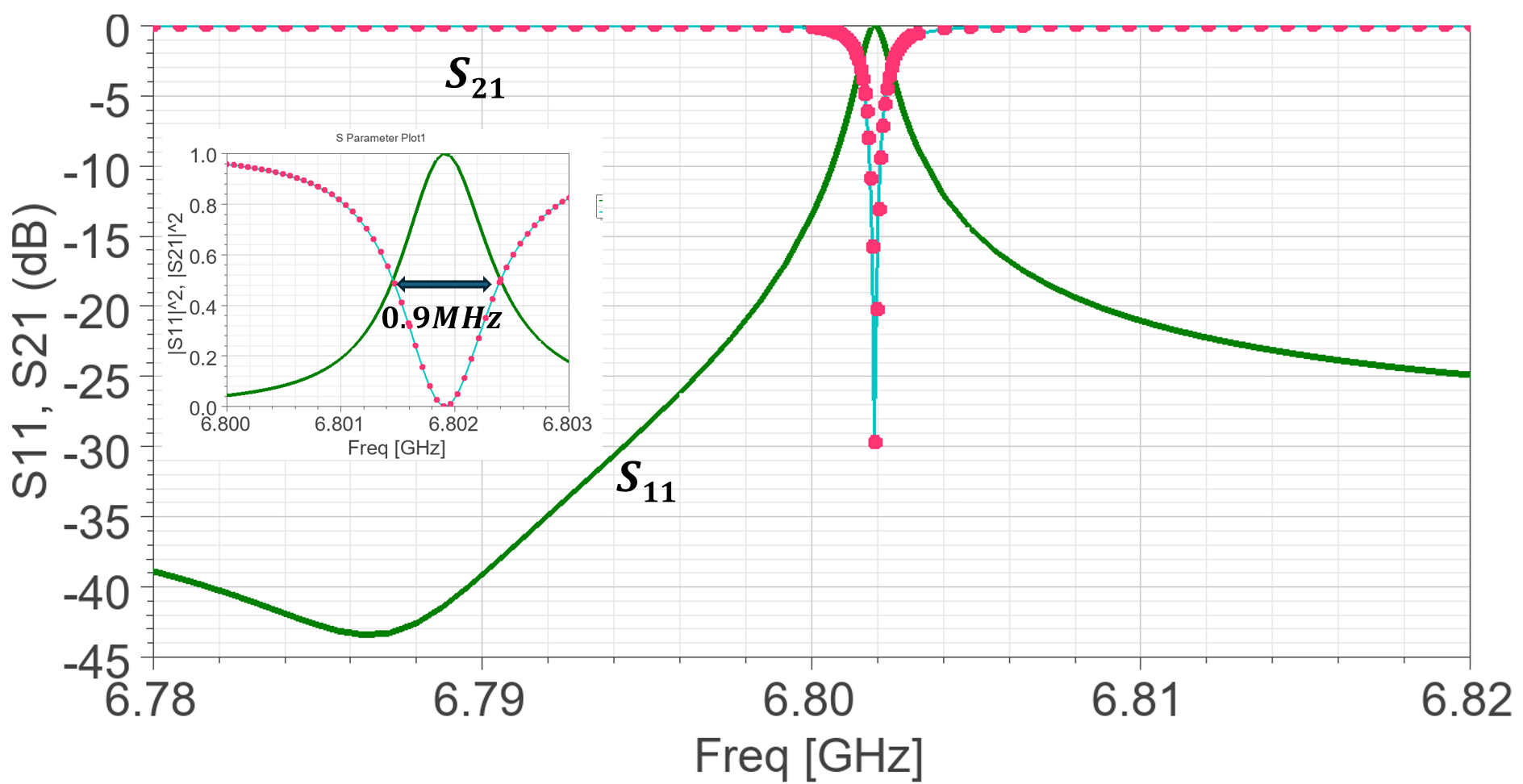}
	\caption{Simulated $S_{11}$ and $S_{21}$ (in dB) of the coupled feedline-resonator system. The inset zooms in on the dip and plots $|S_{11}|^2$ and $|S_{21}|^2$. The BW measured at 50\% of power transferred is indicated and is about 0.9 MHz.}
    \label{fig:S21-TL-Res-dip}
\end{figure}

Fig.~\ref{fig:S21-TL-Res-dip} shows the simulated $S_{11}$ and $S_{21}$ (in dB) of the coupled feedline-resonator system. The inset zooms in on the dip and plots $|S_{11}|^2$ and $|S_{21}|^2$. The $BW$ measured at 50\% of power transferred is indicated, following the definition in Section \ref{Quality_factor_section}. The $BW$ is consistent with the $Q$ extracted in eigenmode simulation through Eq.~(\ref{eq:Q-definition_BW}). 

\subsection{Suggested Exercises}

\begin{itemize}
  \item Try different lengths of the transmission line and plot the eigenfrequencies as a function of length. Does it behave as expected?
  \item The coupling strength between the feedline and the resonator depends on their distance and overlapping length. Design systems with different distances and overlapping lengths between the feedline and the resonator. Does $Q$ change as expected?
\end{itemize}

\section{Transmon Qubit Design} \label{sec:transmon_qubit}

\subsection{Theory}
\begin{figure}[H]
	\centering
	\includegraphics[scale=0.35]{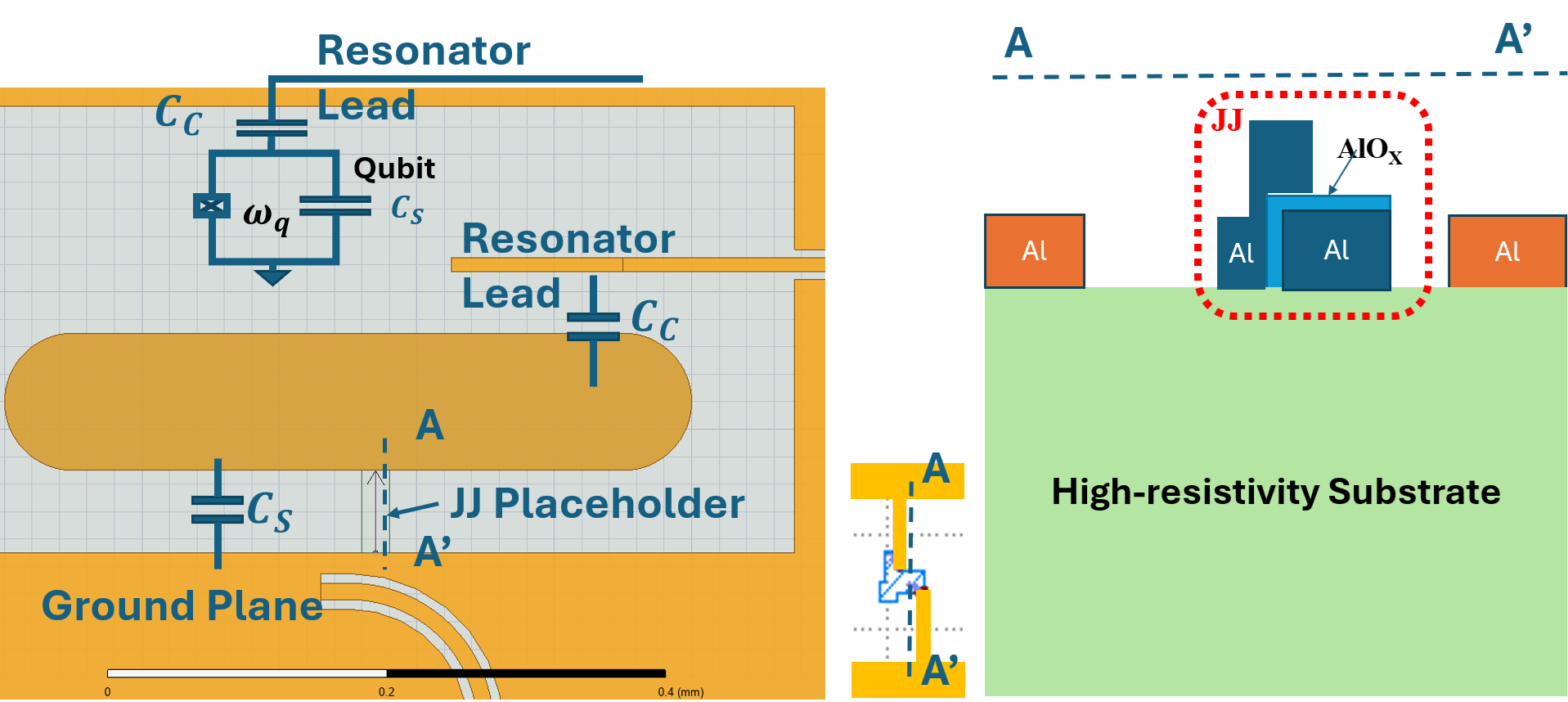}
	\caption{Left: Layout of a transmon coupled to a resonator. The corresponding circuit diagram is shown at the top. Note that the JJ layout is not shown (only has a placeholder). Middle: Zooming in on the JJ layout region. Right: Cross-section of the structure along A-A'. The cross-section of the JJ after fabrication is shown in the middle. Note that the microwave circuit may use materials (orange) different from those of the JJ electrodes.}
    \label{fig:Transmon_Cross_Section}
\end{figure}

A transmon is a type of superconducting qubit that has been widely used in the industry because it is resilient to charge noise \cite{wong2025Transmon}\cite{koch2007}. It can be treated as an LC oscillator where the inductance is given by a Josephson junction (JJ) instead of a regular linear inductor (Fig.~\ref{fig:Transmon_Cross_Section}). The JJ can provide a non-linear inductance, which is important to avoid state leakage and will be discussed below.

\subsubsection{Josephson Junction}
A Josephson junction (JJ) is just a superconductor/insulator/superconductor (SIS) stack (right of Fig.~\ref{fig:Transmon_Cross_Section}). There are various types of JJ \cite{Wong_review}. The common material for the superconductor is aluminum, and the insulator is aluminum oxide. Al becomes superconducting below 1.2 K. JJs obey the two JJ equations,
\begin{equation}
\begin{cases}
I(t)&= I_c \sin\varphi(t), \\
 \frac{\partial \varphi(t)}{\partial t} &=\frac{2\pi }{\Phi_0} V,
\end{cases}
\label{eq:JJ_Equations}
\end{equation}
where $I(t)$ is the current through the JJ at time $t$, $I_c$ is the critical current, $\Phi_0$ is the flux quantum, $V$ is the voltage across the JJ, and $\varphi(t)$ is the quantum phase across the JJ (Josephson phase). Note $\Phi_0=\frac{h}{2e}$ where $h$ is the Planck constant and $e$ is the elementary charge. It can be seen that even if $V=0$, as long as $\varphi(t)\neq 0$, there is a persistent current, which is expected for a superconducting system. When $V$ is a constant, the Josephson phase will change linearly, and thus the current will oscillate under a constant voltage.

Using the inductor equation, $V = L\frac{dI}{dt}$, and expressing it in terms of the Josephson phase, one will find that it has an effective inductance given by \cite{wong2025JJ},

\begin{align}
       L  &=\frac{1}{I_c \cos \varphi(t) \frac{2\pi }{\Phi_0}},\nonumber\\
        &= L_J \frac{1}{\cos \varphi(t)},
\end{align}
where $L_J$ is the Josephson inductance. $L$ is not a constant and depends on $\cos \varphi(t)$. Therefore, the inductance is nonlinear. It should also be noted that, on an integrated circuit, to obtain an inductance on the order of 10 nH, it requires a large regular inductor (e.g., spiral coil), which occupies a very large footprint. However, a JJ can provide $L=10$ nH easily with a footprint in less than a few \textmu m including leads (middle of Fig.~\ref{fig:Transmon_Cross_Section}). And the area of a JJ is only about 0.1 \textmu m $\times$ 0.1 \textmu m. This is because a regular inductor relies on the magnetic field to store the energy (the geometric inductance) but a JJ stores its energy in the Josephson coupling energy, which depends on the superconducting phase difference across the junction. The energy stored by $L_J$ (the Josephson energy) is
\begin{align}
       E_J &= \frac{I_c\Phi_0}{2\pi},\nonumber\\
       &=  \frac{\Phi_0^2}{(2\pi)^2 L_J}.
       \label{eq:JJEnergy}
\end{align}

\begin{figure}[H]
	\centering
	\includegraphics[scale=0.45]{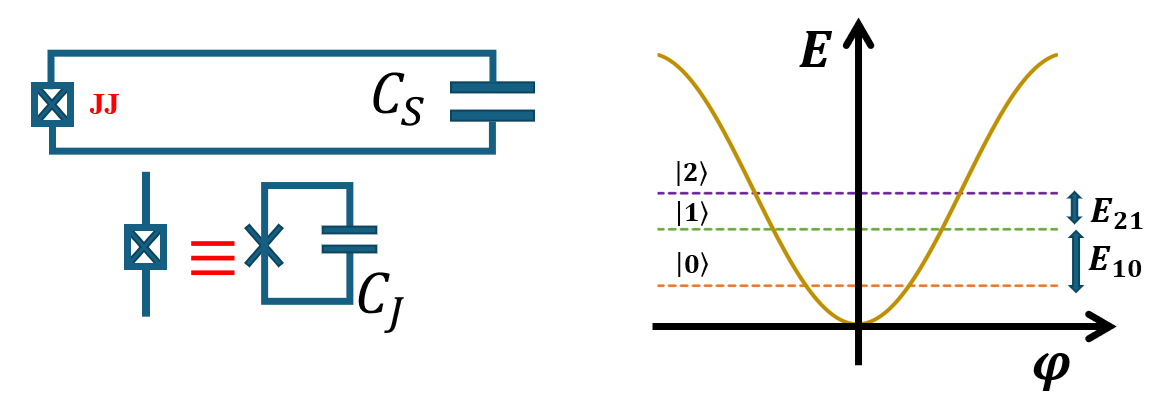}
	\caption{Left: Equivalent circuit of a transmon. $C_J$ is the JJ intrinsic capacitance and $C_S$ is the shunt capacitance. Right: Energy quantization of a transmon versus phase $\varphi$.}
    \label{fig:Transmon_energy}
\end{figure}

Since a JJ is just an SIS stack, it has an intrinsic capacitance ($C_J$) (Fig.~\ref{fig:Transmon_energy}). This is a few fF. 

Moreover, the insulator is thin to allow a weak coupling between the electron wavefunction in one electrode and the other. At room temperature, there is a finite leakage current due to quantum mechanical tunneling. Therefore, it has an effective resistance at room temperature, the normal state resistance $R_n$, given by 
\begin{equation}
R_n=\frac{\pi\Delta}{2I_ce},
\label{eq:Rn_equation}
\end{equation}
where $\Delta$ is the superconducting gap and is $2.92\times 10^{-23}~\text{J}$ for Al. In a foundry process, there is a limit on the $R_n$ range that can be provided, and chip design must adhere to this range.

\subsubsection{Transmon Qubit Energy}

By combining a shunt capacitor ($C_S$) with a JJ, a transmon qubit is formed (Fig.~\ref{fig:Transmon_energy}). This is just an LC oscillator with a non-linear inductance. Therefore, it also has a corresponding resonant frequency, the qubit frequency $f_q$ (or angular frequency, $\omega_q = 2\pi f_q$). Quantum mechanically, the amount of energy it can store is quantized (Fig.~\ref{fig:Transmon_energy}). The energy difference between the first excited state ($\ket{1}$) and the lowest state ($\ket{0}$) is $E_{10}$ and it is related to the qubit frequency by
\begin{equation}
       E_{10}  =\hbar \omega_q,
\end{equation}
where $\hbar=\frac{h}{2\pi}$ is the reduced Planck constant.

Since this is an LC oscillator, it can be modeled as a lumped capacitor ($C_\Sigma$) and a lumped inductance, $L_J$. $C_\Sigma=C_S+C_J+C_{others}$, where $C_{others}$ is the effective total parasitic capacitance across the transmon. One can define the capacitive energy $E_C$ as
\begin{eqnarray}
E_C = \frac{1}{2}\frac{e^2}{C_\Sigma},
\label{eq:charging_energy}
\end{eqnarray}
which is also called the charging energy of a single electron, which is the energy required to add one electron to the capacitor. Therefore, approximately, the qubit angular frequency should be proportional to $1/\sqrt{L_J C_{\Sigma}}$. Using Eqs.~(\ref{eq:JJEnergy}) and (\ref{eq:charging_energy}), $\omega_q\approx1/\sqrt{L_J C_{\Sigma}}$, and $\Phi_0=\frac{h}{2e}$, one can obtain, $\hbar\omega_q\approx\sqrt{8E_CE_J}$. However, since the inductance is non-linear, the qubit energy is given by \cite{wong2025Transmon},
\begin{equation}
       E_{10} =\hbar \omega_q =\sqrt{8E_C E_J}-E_C.     \label{eq:Transmonenergy_Ec}
\end{equation}

Fig.~\ref{fig:Transmon_energy} shows that the transmon qubit may have higher energy levels. However, the spacing between adjacent energy levels is not constant. This is due to the nonlinearity of the Josephson inductance. Therefore, it is not a simple harmonic oscillator, where the energy spacing is constant. The deviation from a perfect harmonic oscillator is called the anharmonicity, $\alpha$, which is defined as and related to $E_C$ through
\begin{equation}
       \alpha =E_{21}-E_{10}\approx-E_C.   \label{eq:anharmonicity}
\end{equation}
Note that $E_{21}$ is smaller than $E_{10}$ in a transmon. Therefore, $\alpha<0$. 
Anharmonicity is important to keep the qubit in the lowest two levels. Without anharmonicity, the same microwave pulse switching the qubit states between $\ket{0}$ and $\ket{1}$ may bring it to other states. If the state leaves the $\ket{0}$ and $\ket{1}$ space, it is called a state leakage. Therefore, $E_C$ needs to be sufficiently large to avoid state leakage.

\subsubsection{Transmon Qubit Design Criteria} \label{criteriasection}
To design a transmon qubit, we need to first pick a qubit frequency. It is usually between $3$ and $6~\text{GHz}$. To operate in the transmon regime, we target 
\begin{equation}
       \frac{E_J}{E_C}>50.   \label{eq:EJEC}
\end{equation}

To have enough anharmonicity, we need
\begin{equation}     E_C>h\times100~\text{MHz}.\label{eq:EC_criteria}
\end{equation}

Finally, $R_n$ should fall in the range of the available values for a given fabrication process with
\begin{equation}     R_{\mathrm{n,min}}<R_n<R_{\mathrm{n,max}}.\label{eq:Rn_criteria}
\end{equation}

\subsection{Simulation}
\label{simulation_transmon_resonator}
From the previous Section, it can be seen that the qubit frequency and anharmonicity depend on the effective capacitance and the Josephson inductance. Here, we will explain how to perform capacitance extraction and qubit eigenmode simulations.

\subsubsection{Capacitance Extraction}

The transmon layout is shown in Fig.~\ref{fig:Transmon_Cross_Section}. The layout generation deck can be found in \cite{GitHub_Capacitance_Extraction} and the corresponding video can be found in \cite{YouTube_Capacitance_Q3D}. In Fig.~\ref{fig:Transmon_Cross_Section}, we can see that there are 2 capacitances, namely, $C_S$, and $C_C$, where $C_C$ is the coupling capacitance to the resonator. Our goal is to extract these 2 capacitances.

We start by opening a project with a single transmon - resonator - CPW transmission line system, and importing the project into a Q3D environment. First, we assign a \textit{thin conductor} to the elements we're interested in, highlighted as orange in Fig.~\ref{fig:Transmon_Cross_Section}. We assign a \textit{PEC} (Perfect Electric Conductor) boundary to each highlighted element and assume they have a thickness of $200\text{ nm}$. 

Next, we assign nets to the connected elements and name them accordingly. The transmon capacitive plate, ground, and resonator lead are called \textit{Transmon}, \textit{GND}, and \textit{Resonator}, respectively.

Table  \ref{tbl:q3d_capacitance_setup} summarizes the numerical setup.
\begin{table}[H]
\centering
\refstepcounter{table}
\addcontentsline{lot}{table}{Table \thetable\quad Q3D Capacitance Simulation Setup}
Table \thetable\\[0.5ex]
Q3D Capacitance Simulation Setup\\[1.5ex]
\label{tbl:q3d_capacitance_setup}

\renewcommand{\arraystretch}{1.3}

\begin{tabular}{|l|l|}
\hline
\textbf{Parameter} & \textbf{Value} \\
\hline
Solution Frequency & 1 GHz \\
\hline
Capacitance/Conductance & True \\
\hline
Automatically Increase Accuracy & True \\
\hline
Accuracy Level & High \\
\hline
Iterative Solver & True \\
\hline
Maximum Number of Passes & 20 \\
\hline
Minimum Number of Passes & 10 \\
\hline
Minimum Converged Passes & 1 \\
\hline
Percent Error & 0.5\% \\
\hline
Percent Refinement per Pass & 30\% \\
\hline
\end{tabular}
\end{table}

The results are summarized by a capacitance matrix (Table  \ref{tbl:cap_matrix_ff}) containing the self capacitances in the diagonal elements, and mutual capacitances in the off-diagonal elements. 

\begin{table}[H]
\centering
\refstepcounter{table}
\addcontentsline{lot}{table}{Table \thetable\quad Capacitance Matrix Between Nets (fF)}
Table \thetable\\[0.5ex]
Capacitance Matrix (fF)\\[1.5ex]
\label{tbl:cap_matrix_ff}

\renewcommand{\arraystretch}{1.3}

\begin{tabular}{|l|c|c|c|}
\hline
\textbf{} & \textbf{GND} & \textbf{Resonator} & \textbf{Transmon} \\
\hline
\textbf{GND} & 881.7 & -663.5 & -82.4 \\
\hline
\textbf{Resonator} & -663.5 & 673.2 & -8.6 \\
\hline
\textbf{Transmon} & -82.4 & -8.6 & 92.7 \\
\hline
\end{tabular}
\end{table}
Therefore, $C_C=8.6~\text{fF}$ and $C_S=82.4~\text{fF}$. Here we assume that the JJ intrinsic capacitance is $C_J=8~\text{fF}$. Then $C_\Sigma=(8.6+82.4+8)~\text{fF}=99~\text{fF}$. Based on Eqs.~(\ref{eq:charging_energy}) and (\ref{eq:anharmonicity}), the anharmonicity of the transmon is
\begin{align}
    \alpha/h&=- 
\frac{e^{2}}{2C_\Sigma} \Big{/} h,\nonumber\\
    &=-\frac{(1.602 \cdot 10^{-19} \text{C})^{2}}{2 \cdot (99 \cdot 10^{-15}\text{F})} \Big{/} (6.626 \cdot 10^{-34} \text{Js}), \nonumber \\ 
    &= -195 \text{ MHz}.
    \label{eq:anharmonicity_calculaton}
\end{align}
The value meets the criterion that $E_C>h\times100~\text{MHz}$.

\subsubsection{Qubit Eigenmode Simulation and Josephson Inductance} \label{sec:1qubiteigenmode}

HFSS is a classical EM simulator. It cannot simulate the nonlinearity of Josephson inductance. To model the Josephson inductance ($L_J$) and the intrinsic capacitance ($C_J$), one can set a lumped RLC boundary for the transmon in the "JJ placeholder" region in Fig.~\ref{fig:Transmon_Cross_Section}. The Jupyter notebook to generate the structure and the video demonstrating how to set up the lumped RLC boundaries can be found in \cite{GitHub_EPR_Extraction} and \cite{YouTube_EPR_CrossKerr}, respectively. One can then perform an eigenmode simulation for the full circuit using the method demonstrated in Section~\ref{sec: tl-res eigenmode}. We set $C_J=8~\text{fF}$ and $L_J=11.1~\text{nH}$. Using Eq.~(\ref{eq:Transmonenergy_Ec}) and the extracted capacitance in the previous Section, $f_q=4.61~\text{GHz}$. The simulated qubit frequency is found to be $f_q=4.50~\text{GHz}$. However, it should be noted that HFSS treats the JJ inductance as a linear inductance. Therefore, the value we should compare to is the one calculated by $\hbar\omega_q\approx\sqrt{8E_CE_J}$, which will be larger than 4.6 GHz. This means that the effective capacitance in the eigenmode simulation is larger than the one extracted with Q3D. However, if the simulation is correct, the experimental value should be 4.5 GHz $-E_C/h$. Based on the value in Eq.~(\ref{eq:anharmonicity_calculaton}), it is 4.305 GHz. A more accurate prediction will be obtained if we use the anharmonicity extracted from the energy participation ratio (EPR) method, as discussed in Section~\ref{EPR_subsection}.

Moreover, based on Eq.~(\ref{eq:JJEnergy}), $E_J=h\times14.8~\text{GHz}$ and is 76 times $E_C$. Therefore, the design fulfills the criterion in Eq.~(\ref{eq:EJEC}).

\begin{figure}[H]
	\centering
	\includegraphics[scale=0.4]{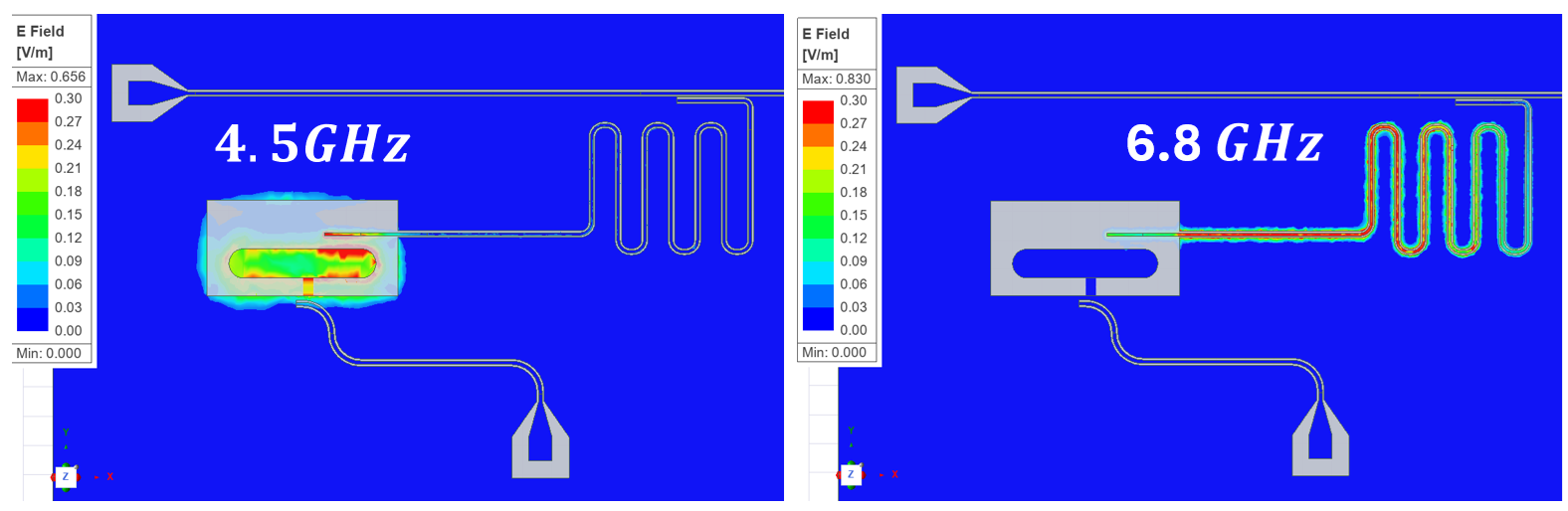}
	\caption{Left: Electric field distribution of Mode 1 (qubit mode). Right: Electric field distribution of Mode 2 (resonator mode).}
\label{fig:Transmon_Eigenmode}
\end{figure}

Fig.~\ref{fig:Transmon_Eigenmode} shows the electric field distribution of the first two modes. The resonator mode (which is the same as that in Fig.~\ref{fig:Eigenmode_Field}) is well separated from the qubit mode (2.3 GHz difference).

One can also add 50 $\Omega$ lumped RLC boundaries to the feedline and redo the eigenmode simulations to obtain the quality factor, $Q_q$, of the qubit, using the method in Section~\ref{sec: tl-res eigenmode}. It is found that $Q_q\approx 2.2\times 10^{6}$. Therefore, the qubit's lifetime, limited by leakage through the coupled resonator and then the feedline, is about 78 $\mu$s. It should be noted that the qubit can also lose its excitation through the flux bias line below it if a lumped RLC boundary is added to the launch pad of the flux bias line. This will result in an even shorter lifetime.

\subsection{Suggested Exercises}
\begin{itemize}
  \item Try to minimize the layout of the transmon but keep $L_J$ unchanged so that it still fulfills the design criteria.
  \item In the simulation, add 50 $\Omega$ lumped RLC boundaries to the feedline as was done in Section~\ref{sec: tl-res eigenmode} to verify $Q_q$ given in the text.
\end{itemize}
\section{Flux Tunable Qubit}
\label{sec:fluxtunable}
\subsection{Theory}
\subsubsection{Flux Tunable Qubit}
\label{sec: DC Bias analytical}

As demonstrated in Section~\ref{sec:transmon_qubit}, the qubit frequency, $f_q$, depends on Josephson inductance, $L_J$, (Eq.~(\ref{eq:Transmonenergy_Ec})). It then further limits the choice of the shunt capacitance in order to fulfill the design criteria in Eqs.~(\ref{eq:EJEC}) to (\ref{eq:EC_criteria}). Therefore, $L_J$ plays a very central role in qubit design. In many qubit operations (such as 2-qubit gates), $f_q$ needs to be tuned over a certain range during the operations. Moreover, due to fabrication variations, sometimes $f_q$ needs to be fine-tuned to the target value. Therefore, it is desirable to have a JJ with electrically tunable $L_J$. A flux-tunable JJ can be used for this purpose.

\begin{figure}[H]
	\centering
	\includegraphics[scale=0.4]{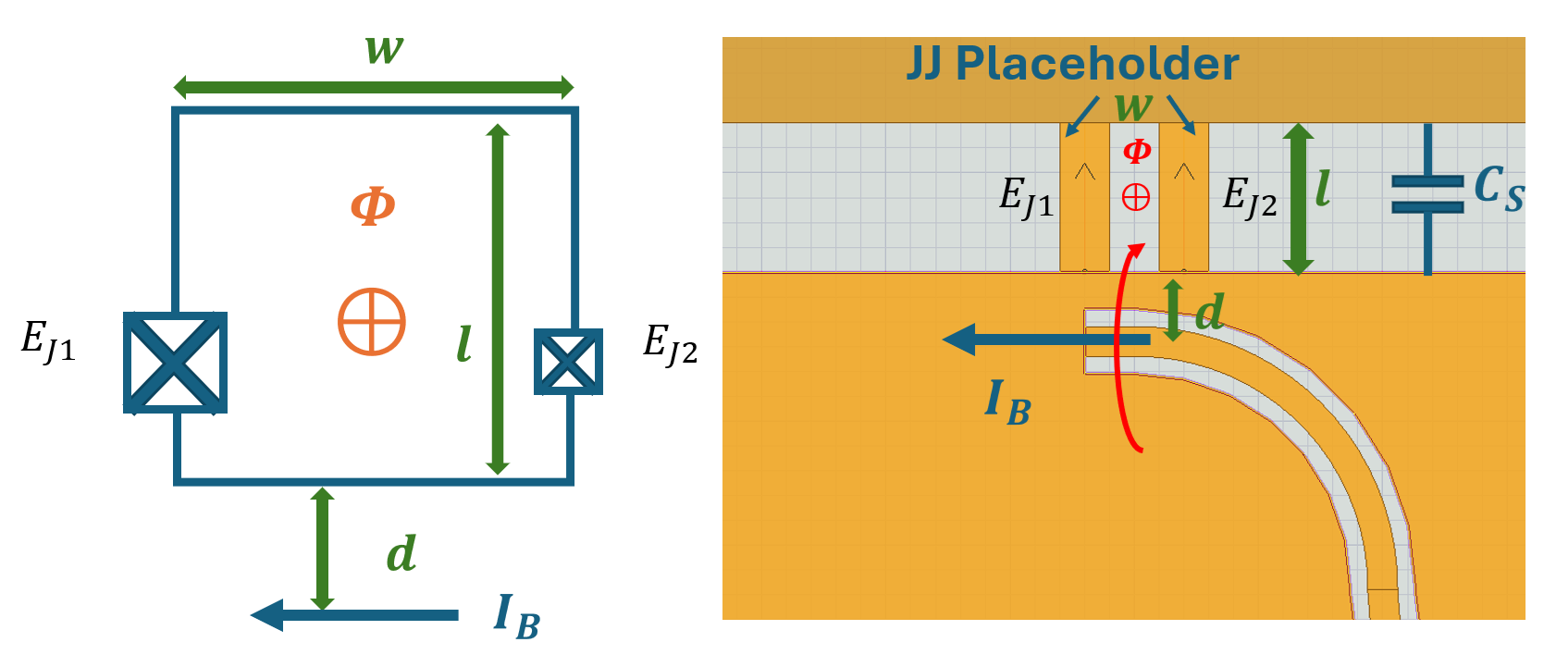}
	\caption{Left: Circuit diagram of a flux-tunable Josephson Junction with a bias current flowing at the bottom from right to left. Right: The layout of a flux-tunable transmon with a flux-tunable JJ loop. The exact layout of the JJs has not been placed. The bias current, $I_B$, for flux generation is included.}
\label{fig:Flux_tunable_qubit}
\end{figure}

Fig.~\ref{fig:Flux_tunable_qubit} shows that a flux-tunable JJ is composed of two JJs connected in a loop. This loop is also called a SQUID loop. SQUID stands for Superconducting Quantum Interference Device. The two JJs need not be identical and thus can have different Josephson energies, $E_{J1}$ and $E_{J2}$. When there is an external flux passing through the loop, the effective $E_J$ will change, and thus it is flux-tunable.

Assuming the two JJs are identical, each with $E_J$, the total Josephson energy is~\cite{wong2025JJ}
\begin{equation}     E_{JT}= 2E_{J}\left| \cos\left(\frac{\pi\Phi}{\Phi_0}\right)\right|,\label{eq:fluxtunableJJ}
\end{equation}
where $\Phi$ is the external flux passing through the loop (Fig.~\ref{fig:Flux_tunable_qubit}). By changing $\Phi$, the effective Josephson energy can be changed from 0 to $2E_J$. That means the effective Josephson inductance can be tuned from $L_J/2$ to infinity, based on Eq.~(\ref{eq:JJEnergy}).

A flux-tunable transmon qubit is usually composed of a flux-tunable JJ loop. On the right of Fig.~\ref{fig:Flux_tunable_qubit}, it shows the layout of a tunable transmon. The tunable JJ loop is in parallel with a fixed shunt capacitor. A bias current, $I_B$, flows through a CPW in the proximity of the JJ loop and generates a circular magnetic field. We need to design it so that the maximum flux threading through the loop due to the magnetic field is large enough to tune $E_{JT}$. On the other hand, we also need to make sure that the current density of $I_B$ is below the critical current that will break the superconductivity of wire.

\subsubsection{Flux Bias Design}
To analytically calculate the flux due to bias current $I_B$, we need to use an ideal setup, which is shown on the left of Fig.~\ref{fig:Flux_tunable_qubit}. We assume that the loop has an area of $w\times l$ and the wire center is at a distance $d$ from the loop. Utilizing Ampere's Law, we get the well-known result for the magnitude of the magnetic field produced by an infinitely long wire at a distance $y$ from the wire~\cite{griffiths2017}:
\begin{equation}
    B = \frac{\mu_{0}I_B}{2\pi y}.
    \label{eq: B field in a wire}
\end{equation}
The total flux threading through the loop is given by integrating the magnetic field over the loop area. Therefore,
\begin{align}
    \Phi &= \frac{\mu_{0} I_B}{2 \pi} \int_{x=0}^{x=w} \int_{y=d}^{y=d + l} \frac{1}{y} \, dy \, dx \nonumber \\
         &= \frac{\mu_{0} I_B}{2\pi} w \, \ln\left( \frac{d + l}{ d} \right).
         \label{eq:flux}
\end{align}
The number of flux quanta (can be non-integer) in the loop $n$ is then given by:

\begin{equation}
    n = \frac{\Phi}{\Phi_{0}}.
    \label{eq: quanta}
\end{equation}
Therefore, the current needed to achieve $n$ flux quanta is given by

\begin{equation}
    I_B= \frac{2\pi \Phi_{0}n}{\mu_{0}} \frac{1}{w \; \text{ln} \Big{(} \frac{d + l}{ d}\Big{)}},
    \label{eq:current_per_quanta}
\end{equation}
It should be noted that this equation is an approximation because the current carrier is not a circular rod but a rectangular thin film, and it is not infinitely long. To be able to fully tune $E_{JT}$, we need to achieve at least half of a flux quantum, i.e., $\Phi_0/2$ (Eq.~(\ref{eq:fluxtunableJJ})). In the layout shown on the right of Fig.~\ref{fig:Flux_tunable_qubit}, the area between the JJ placeholders is $w\times l=$ 20 \textmu m$\times$  60 \textmu m and $d$ = 28 \textmu m. Therefore, to achieve  $\Phi_0/2$, we need 226 \textmu A.

Now assume the thickness of the film is 200 nm. Since the width of the wire is 12 \textmu m, the current density is $9.4\times 10^7$ A/m$^2$, which is less than the critical current density of Al.

\subsection{Simulations}
Now we will perform simulations to find the flux generated by the non-ideal structure on the right of Fig.~\ref{fig:Flux_tunable_qubit}. Maxwell 3D will be used to simulate the flux. The Jupyter notebook for generating the structure and the final HFSS and Maxwell 3D simulation project can be found in~\cite{GitHub_Bias_Flux_Simulation}. The procedure is demonstrated in \cite{YouTube_Bias_Flux_Maxwell}.

\begin{figure}[H]
	\centering
    \includegraphics[scale=0.4]{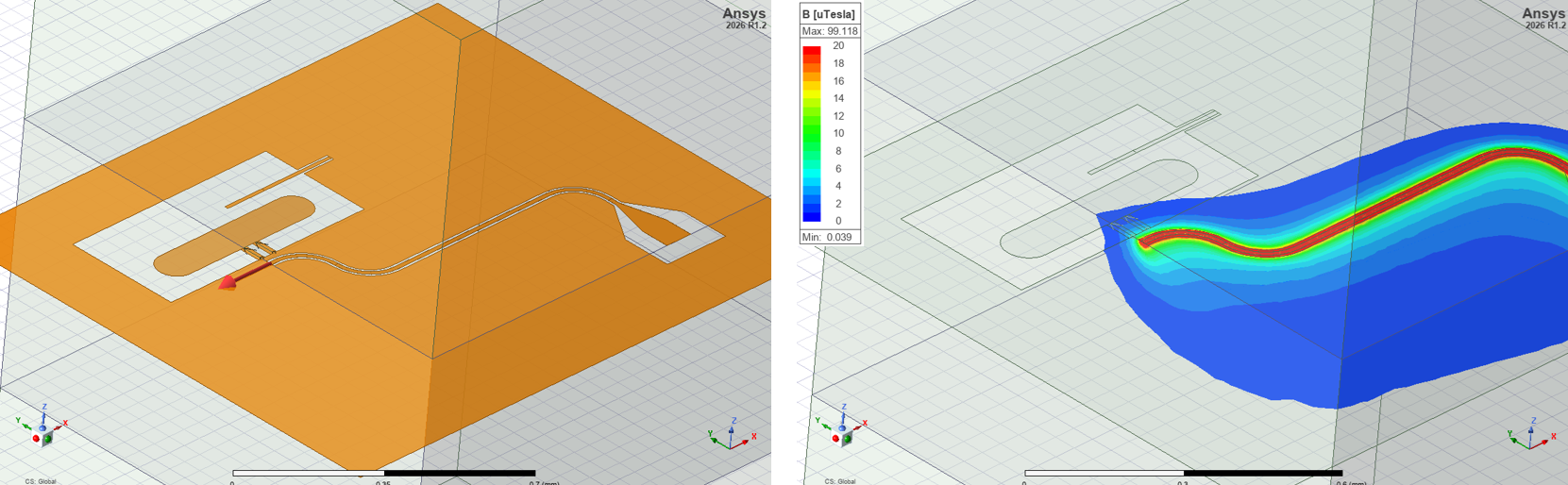}
	\caption{Bias current and magnetic field simulation in Maxwell 3D. Left: The structure with current direction shown (one of the boundary conditions). Right: B-field distribution with $I_B$ = 1 mA.}
    \label{fig: DC bias line rendering}
\end{figure}

After generating the structure in HFSS, we need to export it to Maxwell 3D. Fig.~\ref{fig: DC bias line rendering} shows the structure in Maxwell 3D where it is truncated. This is because we need to cut the substrate and vacuum such that the DC bias line source begins and terminates at the boundary interface.

To set up the boundary conditions in Maxwell 3D, firstly, we click on the conductive region of the CPW, extrude it upward so it has a thickness of $200~\text{nm}$, and assign it a PEC (Perfect Electric Conductor) boundary. We can also delete the vacuum box and silicon substrate, and redraw new ones such that the current boundary condition arrows are aligned with the normal of the boundary box surfaces. One of the boundary condition setups is visualized as the arrow on the left of Fig.~{\ref{fig: DC bias line rendering}}. $I_B=1$ mA is used.

We configure the setup to be 15 maximum number of passes, 8 minimum number of passes, and $1\%$ error. Further setup details are summarized in Table  \ref{tbl:maxwell3d_magnetostatic}.

\begin{table}[H]
\centering
\refstepcounter{table}
\addcontentsline{lot}{table}{Table \thetable\quad Maxwell 3D Magnetostatic Setup}
Table \thetable\\[0.5ex]
Maxwell 3D Magnetostatic Setup\\[1.5ex]
\label{tbl:maxwell3d_magnetostatic}

\begin{tabular}{|l|l|}
\hline
\textbf{Parameter} & \textbf{Value} \\
\hline
Maximum Number of Passes & 15 \\
\hline
Percent Error & 1\% \\
\hline
Solve Fields Only & True \\
\hline
Refinement per Pass & 30\% \\
\hline
Minimum Number of Passes & 8 \\
\hline
Minimum Converged Passes & 1 \\
\hline
\end{tabular}
\end{table}

The right of Fig.~{\ref{fig: DC bias line rendering}} shows the simulation result by plotting the magnetic field. It can be seen that the magnetic field strength decreases with distance from the current source. However, there is still a substantial magnetic field in the SQUID loop.

To calculate the magnetic flux, we draw a non-model rectangle over the region where the flux is to be calculated in the SQUID loop. Then, we use the fields calculator to calculate the amount of magnetic flux through the \textit{flux region}. To open the fields calculator, navigate to \textit{Maxwell 3D} $>>>$ \textit{Fields} $>>>$ \textit{Calculator}. In the calculator, we go through the series of button presses to do a surface integral to calculate the magnetic flux through the \textit{flux region}: \textit{Quantity} $>>>$ \textit{B} $>>>$ \textit{Geometry} $>>>$ \textit{Surface} $>>>$ \textit{flux\_region} $>>>$ \textit{Normal} $>>>$ $\int$ $>>>$ 
\textit{Eval} (follow also \cite{YouTube_Bias_Flux_Maxwell}). The output of the calculator suggests that we have a magnetic flux of $6.37\times10^{-15} \text{ Wb}$. Therefore, the $226~\mu A$ calculated in the last Section is expected to produce $1.44\times10^{-15} \text{ Wb}$ which is $0.7\Phi_{0}$. This is larger than required, and the design is expected to work correctly.

\subsection{Suggested Exercises}
\begin{itemize}
  \item When the flux bias line is closer to the qubit, the qubit is more susceptible to noise. Try to move the bias line away from the qubit and simulate to find the maximum distance so that it can still achieve full-scale tuning when $I_B=2~\mathrm{mA}$.
  \item In the simulation, add a 50 $\Omega$ lumped RLC boundary to the launch pad of the flux bias line to perform eigenmode simulations to find the $Q$ of the qubit.
\end{itemize}

\section{Qubit Readout}
\label{sec:qubitreadout}
\subsection{Theory}
\subsubsection{Dispersive Shift and Cross-Kerr}
The state of a qubit can be measured through various methods. One of the common methods is a dispersive readout scheme~\cite{wong2025readout}. In this scheme, a resonator is coupled to the qubit. Figs.~\ref{fig:Transmon_Cross_Section} and~\ref{fig:Transmon_Eigenmode} implement this scheme. The qubit is capacitively coupled to a resonator through $C_C$. Based on quantum mechanical theory, the resonator frequency ($f_r$) can be affected by the state of the qubit ($\ket{0}$ or $\ket{1}$)~\cite{wong2025readout}.

\begin{figure}[H]
	\centering
    \includegraphics[scale=0.4]{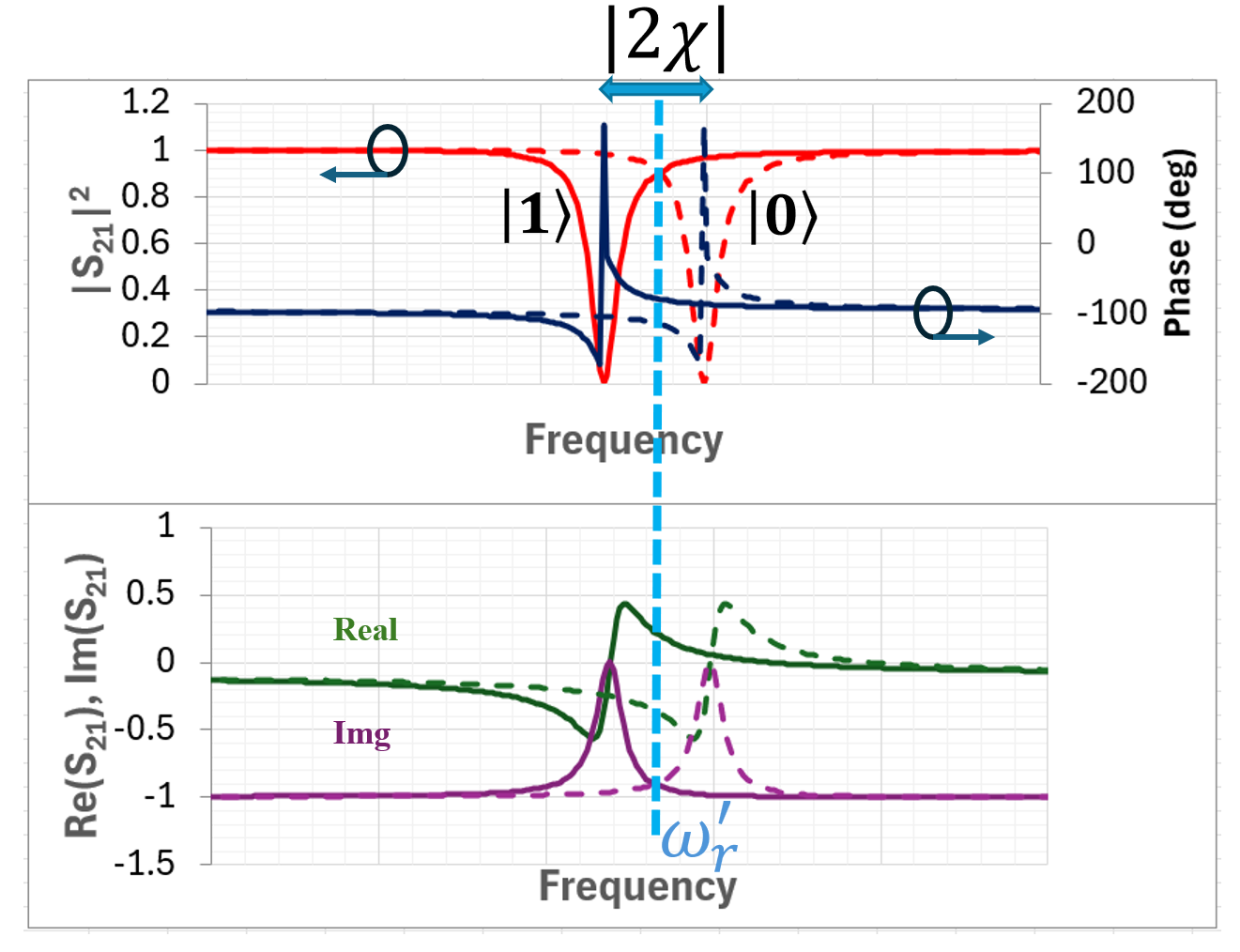}
	\caption{$S_{21}$ of a feedline-resonator system when the resonator is coupled to a qubit (such as the one in Fig.~\ref{fig:Transmon_Cross_Section}). Its magnitude squared and phase are plotted in the top figure, and the real and imaginary parts are plotted in the bottom figure. In this example, $f_r$ when the qubit is at $\ket{0}$ is higher than $f_r$ when the qubit is at $\ket{1}$. Their difference is $|2 \chi|$.}
    \label{fig:Cross-Kerr_theory}
\end{figure}

Fig.~\ref{fig:Cross-Kerr_theory} shows the $S_{21}$ of a feedline-resonator system when the resonator is coupled to a qubit in different states. Let us first discuss what happens when two resonators without anharmonicity are coupled to each other. Assume they have frequencies $\omega_r$ and $\omega_q$, there will be a frequency shift as the modes repel each other. $\omega_r$ will become
\begin{equation}
\omega_{r}''=\omega_r-\frac{|g/\hbar|^2}{\Delta},
\label{eq:frequencyshift}
\end{equation}
where $\Delta=\omega_{q}-\omega_{r}$ is the detuning and $g$ is the coupling strength between the resonators expressed in energy units. This can be understood by the fact that the higher-frequency mode will be pushed higher. $g$ naturally depends on the coupling capacitance between the resonator and the qubit, $C_C$, as shown in Fig.~\ref{fig:Transmon_Cross_Section}.

When a resonator is coupled to a transmon qubit, which is just another resonator with anharmonicity, $\omega_r$ will experience an additional shift due to the anharmonicity depending on the state of the qubit. The difference in the shift due to $\ket{1}$ and $\ket{0}$ is given by $2\chi$, where $\chi$ is
\begin{equation}
\chi =  -\lvert g/\hbar\rvert^2\frac{E_C/\hbar}{\Delta(\Delta-E_C/\hbar)}.
\label{eq:Cross-Kerr}
\end{equation}
For convenience, the midpoint between the two frequencies is often labeled as $\omega_r'$ which has the expression of 
\begin{equation}
\omega_{r}'=\omega_r-\frac{|g/\hbar|^2}{\Delta+\alpha/\hbar}.
\label{eq:frequencyshift_alpha}
\end{equation}
Therefore, the resonator frequency becomes $\omega_r'-\chi$ and $\omega_r'+\chi$ when the qubit state is $\ket{0}$ and $\ket{1}$, respectively.

Eq.~(\ref{eq:Cross-Kerr}) shows that if $\omega_{r}>\omega_{q}$, $\Delta<0$ and $\chi<0$, regardless of the value of $E_C$. Let the shifted resonator frequency be $\omega_{r,\text{shifted}}$. Then, $\omega_{r,\text{shifted}}>\omega_r'$ when the qubit is in state $\ket{0}$. When the qubit is in state $\ket{1}$, $\omega_{r,\text{shifted}}<\omega_r'$. This is shown in Fig.~\ref{fig:Cross-Kerr_theory}. When one measures $S_{21}$ at $\omega_{r}'$ of the feedline, one cannot distinguish whether the qubit is in state $\ket{0}$ or $\ket{1}$ if one only measures its magnitude. However, the phase is different for $\ket{0}$ and $\ket{1}$. Therefore, we can distinguish the state by measuring the phase; equivalently, the real and imaginary parts of $S_{21}$.

In principle, as long as $\chi$ is non-zero, one can distinguish the qubit state. However, due to noise, $\chi$ needs to be large enough \cite{Wong2023ReadoutFidelity}. Therefore, it is important to simulate $\chi$ before the chip is taped out to ensure it has a reasonably large value.

\subsubsection{Energy Participation Ratio Method}

For typical simulators, such as Ansys HFSS, they can only perform classical EM simulations. In order to calculate $\chi$, one may use the energy participation quantization method. This is based on the energy participation ratio (EPR) of the JJ in an electromagnetic mode. Therefore, one only needs to simulate the magnetic field and electric field energy distribution using eigenmode simulation to extract cross-Kerr.

EPR reduces the quantization problem to finding the fraction of the energy of Mode $m$ stored in element $j$, which can be accomplished with HFSS. EPR of Mode $m$ is given by \cite{minev2021}
\begin{equation}
p_m=\frac{\text{Inductive Energy Stored in JJ}}{\text{Total Inductive Energy in the Circuit in Mode $m$}}.
\label{eq:EPR_m}
\end{equation}
When a resonator is coupled to a qubit, the cross-Kerr is given by,
\begin{equation}
2\chi=-p_q p_r\frac{\hbar\omega_q\omega_r}{4E_J},
\label{eq:EPR_m}
\end{equation}
where $p_q$ and $p_r$ are the EPR of the qubit and resonator modes. For example, $p_r$ is the ratio of inductive energy stored in the JJ at the resonator frequency. It should be noted that $\chi_{qc}$ in~\cite{minev2021} is $2\chi$ in this paper. Also, the sign convention of cross-Kerr in this paper is different from that in~\cite{minev2021}.

EPR can be used to find the anharmonicity of the transmon also. It is given by,
\begin{equation}
\alpha=-p_q^2\frac{(\hbar\omega_q)^2}{8E_J}.
\label{eq:EPR_m}
\end{equation}

\subsection{Simulation}
\label{EPR_subsection}

The pyEPR module in Quantum Metal will be used to calculate the cross-Kerr. The Jupyter notebook for generating the structure and the video demonstrating how to use pyEPR can be found at \cite{GitHub_EPR_Extraction} and \cite{YouTube_EPR_CrossKerr}, respectively.

After the structure is generated using Quantum Metal and HFSS, the pyEPR package is used to control the Ansys HFSS project. The code can be used to define the Josephson junction inductance and capacitance. Using the same setup as in Section~\ref{simulation_transmon_resonator}, eigenmode simulations are performed. To speed up the process, Maximum Number of Passes, Maximum Delta Frequency per Pass, and Minimum Converged Passes are set to 15, 1\%, and 1, respectively. $f_q$ and $f_r$ are found to be 4.49 GHz and 6.77 GHz, respectively. They are almost the same as the results in Fig.~\ref{fig:Transmon_Eigenmode}.

Then it will perform EPR analysis, where it will extract the participation ratio, eigen-frequencies, cross-Kerr, and anharmonicity.

Table  \ref{tab:participation_matrix} shows the participation ratio for different modes before and after normalization, where $j1$ represents the only Josephson junction in the circuit.

\begin{table}[H]
\centering
\refstepcounter{table}
\addcontentsline{lot}{table}{Table \thetable\quad Normalized Participation Matrix $P$}
Table \thetable\\[0.5ex]
Participation Matrix before and after Normalization\\[1.5ex]
\label{tab:participation_matrix}

\renewcommand{\arraystretch}{1.3}

\begin{tabular}{|c|c|c|}
\hline
\textbf{Mode $m$} & \textbf{j1} &\textbf{j1} (normalized) \\
\hline
1 &1.041781 &      0.99 \\
\hline
2 &0.003251 &0.0033    \\
\hline
3 & 0.000148 &0.00015     \\
\hline
4 &  0.000247& 0.00025    \\
\hline
5 & 0.000151 & 0.00015  \\
\hline
6 & 0.000034 & 3.4e-05 \\
\hline
\end{tabular}
\end{table}

The result is reasonable. Firstly, $p_{m=1}$ is about 100\%. This is because Mode 1 is the qubit mode ($f_q$) and therefore, the inductive energy concentrates at the JJ. Secondly, the sum of all rows is about 100\% as required by the theory. One can increase accuracy by improving the eigenmode simulation accuracy.

Furthermore, pyEPR produces a $\chi$ matrix, shown in Table~\ref{tab:CrossKerr_matrix} (only the first two modes are shown), that encodes information about the anharmonicity and cross-Kerr between different elements.

\begin{table}[H]
\centering
\refstepcounter{table}
\addcontentsline{lot}{table}{Table \thetable\quad $\chi$ Matrix}
Table \thetable\\[0.5ex]
$\chi$ matrix (MHz). By the convention in pyEPR, the diagonal elements are $-\alpha$ and off-diagonal elements are $-2\chi$.\\[1.5ex]
\label{tab:CrossKerr_matrix}

\renewcommand{\arraystretch}{1.3}

\begin{tabular}{|c|c|c|}
\hline
\textbf{Modes} & \textbf{1}& \textbf{2}\\
\hline
1 &167&  1.71   \\
\hline
2 & 1.71&0.00439      \\
\hline
\end{tabular}
\end{table}
The diagonal elements are the self-Kerr which is the anharmonicity of the mode. Note that in pyEPR, the anharmonicity values are negated. Therefore, the $\alpha$ of the qubit is -167 MHz. This is similar to the calculation in Eq.~(\ref{eq:anharmonicity_calculaton}), which is -195 MHz. The discrepancy again agrees with the postulation that the effective capacitance in eigenmode simulation is larger than the Q3D extraction. The anharmonicity of the resonator mode is only 4.4 kHz and is expected to be small. The off-diagonal elements are $-2\chi$ between the qubit and the resonator. $2\chi=-1.71~\text{MHz}$ is enough for most applications.

\subsection{Suggested Exercises}
\begin{itemize}
  \item The cross-Kerr depends on the coupling strength between the qubit and the resonator, as can be seen in Eq.~(\ref{eq:Cross-Kerr}). Try to vary the layout to change the coupling capacitance $C_C$ and plot $\chi$ as a function of the geometry parameters.
  \item Change the qubit frequency by changing $L_J$. Plot the cross-Kerr as a function of $L_J$. Explain your observation in the $\Delta>0$ and $\Delta<0$ regimes.
\end{itemize}

\section{Three-Qubit Chip Design} \label{sec:3qubit}

\subsection{Theory}

In this last section, we will design a qubit chip with one feedline and three transmons. Two of them are coupled to each other through a half-wavelength resonator. The final layout of the chip is already shown in Fig.~\ref{fig:1_full_chip}. 

\subsubsection{Design Targets}
We pick the qubit and resonator frequencies based on the capability of the measurement equipment and other factors (such as cross-talk minimization). Table \ref{tab:3-qubit_design} shows the desired frequencies. 

\begin{table}[H]
\centering
\refstepcounter{table}
\addcontentsline{lot}{table}{Table \thetable\quad $\chi$ Matrix}
Table \thetable\\[0.5ex]
Design Targets of the 3-Qubit Chip\\[1.5ex]
\label{tab:3-qubit_design}

\renewcommand{\arraystretch}{1.3}

\begin{tabular}{|c|c|c|c|c|c|c|c|}
\hline
\textbf{Modes} & \textbf{Q1}& \textbf{Q2}& \textbf{Q3}& \textbf{R1}& \textbf{R2}& \textbf{R3}& \textbf{RC}\\
\hline
$f$ (GHz) &4.5&  6.22 &  5.81&  6.86&  7.6&  7.4&  5.3  \\
\hline
Length (mm) &N/A&  N/A &  N/A&  4.42&  3.99&  4.1&  11.45 \\
\hline
Total $L_J$ (nH) &11.6&    6.01& 6.87 & N/A&  N/A &  N/A&  N/A \\
\hline
Each $L_J$ (nH) &23.2&    12.0& 13.7 & N/A&  N/A &  N/A&  N/A \\
\hline
\end{tabular}
\end{table}
\subsubsection{Half-wavelength Resonator Coupler}\label{sec:haflwavelength}
A half-wavelength resonator (with both ends open), RC, capacitively coupled to the two top qubits is used for qubit coupling. Similar to the equation for a quarter-wavelength resonator (Eq.~(\ref{eq: quarter-wavelength resonator frequency spectrum})), the resonant frequency of a half-wavelength resonator is,
\begin{equation}
    f = \frac{c}{\sqrt{\epsilon_\mathrm{eff}}\lambda}
    =\frac{cn}{2l\sqrt{\epsilon_\mathrm{eff}}},
\label{eq:halfwavelength}
\end{equation}
where $n$ can be any positive integer and $\lambda$ is the wavelength in the resonator. We design the resonator frequency to be below the coupled qubits (Q2 and Q3). It is set to 5.3 GHz, then the length of the resonator will be 11.45 mm with $\epsilon_\mathrm{eff}=6.1$ (same as Section~\ref{sec: tl-res eigenmode}). 

Table \ref{tab:3-qubit_design} shows the corresponding resonator lengths using Eqs.~(\ref{eq: quarter-wavelength resonator frequency spectrum}) and (\ref{eq:halfwavelength}).

\subsubsection{Josephson Junction and Transmon Design}
In this design, flux-tunable qubits are used. Therefore, two identical JJs will be placed in a SQUID loop in each transmon. The effective total Josephson inductance is thus half of each individual Josephson inductance. One needs to make sure each of them can be fabricated in the chosen foundry process. The total and individual Josephson inductance for achieving the target transmon frequencies are also shown in Table \ref{tab:3-qubit_design}. However, in the simulation and in the example, we only use one JJ with the effective inductance. This simplifies the simulation setup and EPR data analysis.

Note that each single JJ is assumed to have a $C_J=4~\text{fF}$ (instead of $8~\text{fF}$). Therefore, the total effective $C_J$ is still $8~\text{fF}$. For the coupled transmons, they also have extra capacitance due to the coupling to the coupling resonator. They are found to be about $3.7~\text{fF}$ and are included in $C_\Sigma$ for qubit frequency estimation.

\subsection{Simulation}

The Jupyter notebook and HFSS projects for this section can be found in ~\cite{GitHub_3_Qubit_Chip}. The corresponding YouTube video can be found in ~\cite{YouTube_3_Qubit_Chip}.

\subsubsection{Eigenmode}
\begin{figure}[H]
\centering
\includegraphics[width=0.8\textwidth]{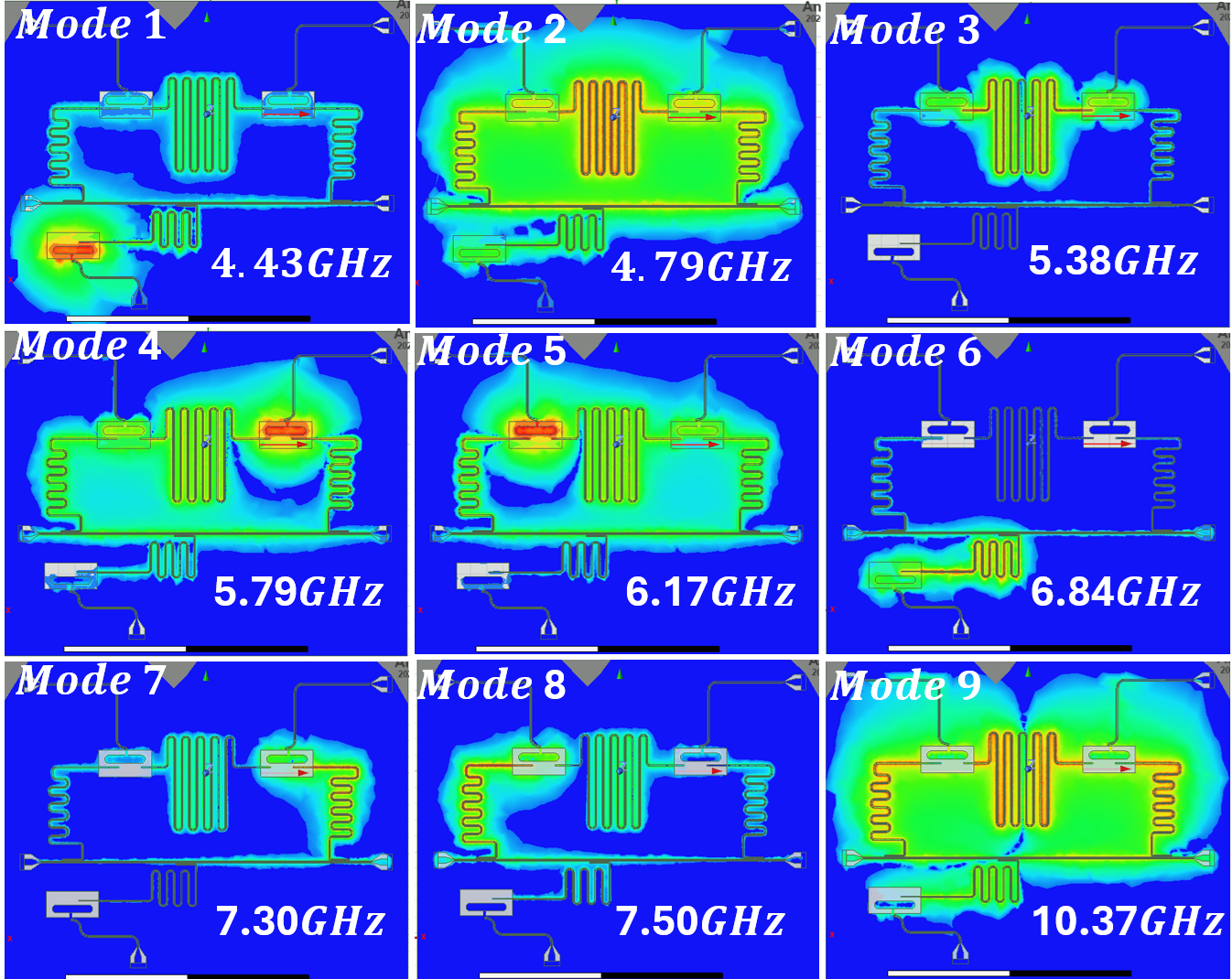}
\caption{E-field distribution at different modes and the mode frequencies of the 3-qubit chip. The structure is shown in Fig.~\ref{fig:1_full_chip}.}
\label{fig:1_full_chip_mode}
\end{figure}

The eigenmode simulation result is shown in Table \ref{tab:3-qubit_design_eigenmode_sim} and Fig.~\ref{fig:1_full_chip_mode}. Note that a $50~\Omega$ lumped RLC boundary is added to the feedline and flux bias line launch pads. Compared to the design in Table \ref{tab:3-qubit_design}, they match pretty well, and most of the deviations are within 0.1 GHz. More importantly, the spacing between the qubits (Q2 vs. Q3) or resonators (R2 vs. R3) is very close to the designed values. However, as discussed in Section~\ref{sec:1qubiteigenmode}, the simulated qubit frequencies miss the $-E_C$ term. By taking that into account, we obtain the last row in Table \ref{tab:3-qubit_design_eigenmode_sim}. These are the expected frequencies to be measured in the experiment. The $E_C$ value for each qubit is obtained in the next section through EPR calculation. Therefore, if the simulation is accurate, the measured qubit frequencies are about 0.2 GHz below the targets.

\begin{table}[H]
\centering
\refstepcounter{table}
\addcontentsline{lot}{table}{Table \thetable\quad $\chi$ Matrix}
Table \thetable\\[0.5ex]
Eigenmode Simulation Results\\[1.5ex]
\label{tab:3-qubit_design_eigenmode_sim}

\renewcommand{\arraystretch}{1.3}

\begin{tabular}{|c|c|c|c|c|c|c|c|}
\hline
\textbf{Modes} & \textbf{Q1}& \textbf{Q2}& \textbf{Q3}& \textbf{R1}& \textbf{R2}& \textbf{R3}& \textbf{RC}\\
\hline
Target $f$ (GHz) &4.5&  6.22 &  5.81&  6.86&  7.6&  7.4&  5.3  \\
\hline
Simulated $f$ (GHz) &4.43&  6.17 &  5.79&  6.84&  7.50&  7.30&  5.38  \\
\hline
Q &5.1e6&  108829 &  119474&  7350.3&  5627.55&  6014.33&  3.15e7 \\
\hline
Lifetime ($\mu$s) &185.24
&  2.81
 &  3.29
&  0.17
&  0.12
&  0.13
&  933.20 \\
\hline
Simulated $f$ with $-E_C$ correction (GHz) &4.26&  6.01 &  5.63&  N/A&  N/A&  N/A&  N/A  \\
\hline
\end{tabular}
\end{table}

There are also two spurious modes, Mode 2 and Mode 9. Both Mode 2 and Mode 9 have substantial energy in the Q2 and Q3 region. This indicates that they might substantially reduce the lifetimes of Q2 and Q3. They might be due to the ground plane island mode above the feedline. This ground plane is fairly isolated in a loop surrounded by the feedline, R2, Q2, RC, Q3, and R3. To remove these modes, one possible solution is to add airbridges or wire bonds across the feedline and resonator lines to tie inner ground to the outer ground \cite{koster1989}\cite{Dhillon_import}. This is left as an exercise.

Using Eq.~(\ref{eq:Q-definition}), one can calculate the lifetime of the qubits and resonators. While Q1 has a reasonable lifetime (leakage through the feedline and the flux bias line), Q2 and Q3 have low lifetimes. One of the reasons is a much stronger coupling between the qubit and the corresponding resonator. Another reason is their potential coupling with Mode 2 and Mode 9, which will be shown to be lossy (to be discussed in the EPR section).

The resonators all have short enough lifetimes to allow readout times of a few hundred ns.

\subsubsection{EPR Calculations}
We perform EPR analysis as discussed in Section~\ref{EPR_subsection}. Table~\ref{tab:participation_matrix_3qubit} shows the normalized EPR for the 9 modes simulated. We can see that Modes 1, 4, and 5 have a larger EPR because those modes correspond to the qubit modes. For example, Mode 4 has an EPR of 0.94. This means that $p_{m=4}$ at $j3$ is large and, indeed, Mode 4 is the qubit mode of Q3 as shown in Fig.~\ref{fig:1_full_chip_mode}.

\begin{table}[H]
\centering
\refstepcounter{table}
\addcontentsline{lot}{table}{Table \thetable\quad Normalized Participation Matrix $P$}
Table \thetable\\[0.5ex]
Normalized Energy Participation Matrix for the 3-qubit Chip\\[1.5ex]
\label{tab:participation_matrix_3qubit}

\renewcommand{\arraystretch}{1.3}

\begin{tabular}{|c|c|c|c|}
\hline
\textbf{Mode $m$} & \textbf{Q1, j1} &\textbf{Q2, j2} &\textbf{Q3, j3}  \\
\hline
1 &\textbf{0.99}	&3.60E-06&	4.80E-06
  \\
\hline
2 &0.00063	&	0.011	&	0.018
     \\
\hline
3 & 3.40E-09	&	0.0024	&	0.0069
      \\
\hline
4 & 1.70E-06	&	0.0057	&	\textbf{0.94}
    \\
\hline
5 & 5.10E-07	&	\textbf{0.94}	&	0.0043
   \\
\hline
6 & 0.0026	&	4.50E-09	&	4.90E-07
  \\
\hline
7 & 3.90E-08	&	7.20E-05	&	0.0096

  \\
\hline
8 & 1.10E-08	&	0.014	&	3.80E-05

  \\
\hline
9 & 9.40E-06	&	0.0047	&	0.0039

  \\
\hline
\end{tabular}
\end{table}

Table~\ref{tab:Chi_3qubit} shows the $\chi$ matrix of the chip. We have the following observations. Firstly, among the diagonal elements, only the modes due to Q1, Q3, and Q2 have significant values. This is because they are the qubit modes. The anharmonicities of Q1, Q3, and Q2 are -170 MHz, -156 MHz, and -156 MHz, respectively (note again pyEPR has an opposite sign to this paper). This fulfills the design requirements in Section~\ref{criteriasection}.

Secondly, $|2\chi|$ between Q1 and R1 is 1.4 MHz. $|2\chi|$ of Q2/R2 and Q3/R3 are much larger (5.55 MHz and 4.01 MHz). They are too high compared to the resonator linewidth. This may be one of the reasons for the shorter lifetime of Q2 and Q3. 

Thirdly, there is a substantial $|2\chi|$ value between Q2 and Q3 (3.3 MHz). This means the Q3 frequency will depend on the state of Q2 and vice versa. This is known as ZZ coupling \cite{kandala2021}. This is likely because Q2 and Q3 are coupled through RC and the spurious modes. If the coupling through RC dominates, this can be alleviated by reducing the coupling strength between RC and Q2/Q3, at the cost of slower two-qubit gates. A tunable coupler can also be used to remedy the effect by switching off the coupling when no two-qubit gate is being performed \cite{sung2021}. If the cause is due to the coupling through the spurious modes, the spurious modes need to be removed by adding airbridges to connect the inner ground plane to the outer ground plane.

Finally, $|2\chi|$ between Q2 (Q3) and Mode 2 is $\sim$3 MHz ($\sim$5 MHz). That between Q2 (Q3) and Mode 9 is $\sim$2 MHz ($\sim$2.4 MHz). Moreover, Modes 2 and 9 have Q of about 2000 and 1000, respectively. These values are not listed in Table~\ref{tab:Chi_3qubit}. This further indicates that the isolated ground plane reduces the lifetime of Q2 and Q3.

\begin{table}[H]
\centering
\refstepcounter{table}
\addcontentsline{lot}{table}{Table \thetable\quad Normalized Participation Matrix $P$}
Table \thetable\\[0.5ex]
$\chi$ Matrix of the 3-qubit Chip. By the convention in pyEPR, the diagonal elements are $-\alpha$ and off-diagonal elements are $-2\chi$\\[1.5ex]
\label{tab:Chi_3qubit}

\renewcommand{\arraystretch}{1.3}

\begin{tabular}{|c|c|c|c|c|c|c|c|}
\hline
$\chi_{i,j}$ (MHz)& \textbf{Q1} &\textbf{RC} &\textbf{Q3} &\textbf{Q2}  &\textbf{R1} &\textbf{R3} &\textbf{R2} \\
\hline
\textbf{Q1} &\textbf{170}	&	1.16E-05	&	0.00198	&	0.00111	&	\textbf{1.4}	&	3.77E-05	&	2.16E-05

  \\
\hline
\textbf{RC} &1.16E-05	&	0.00807	&	2.14	&	0.688	&	1.31E-06	&	0.0275	&	0.0122

     \\
\hline
\textbf{Q3} & 0.00198	&	2.14	&	\textbf{156}	&	3.3	&	0.000193	&	\textbf{4.01}	&	0.0475

      \\
\hline
\textbf{Q2} & 0.00111	&	0.688	&	3.3	&	\textbf{156}	&	3.59E-06	&	0.0478	&	\textbf{5.55}

    \\
\hline
\textbf{R1} & 1.4	&	1.31E-06	&	0.000193	&	3.59E-06	&	0.00287	&	2.54E-06	&	6.45E-08

   \\
\hline
\textbf{R3} & 3.77E-05	&	0.0275	&	4.01	&	0.0478	&	2.54E-06	&	0.0257	&	0.000707

  \\
\hline
\textbf{R2} & 2.16E-05	&	0.0122	&	0.0475	&	5.55	&	6.45E-08	&	0.000707	&	0.0494

  \\
\hline
\end{tabular}
\end{table}

\subsection{Suggested Exercises}
\begin{itemize}
  \item Change the target frequency of Q2 and Q3 to below 5 GHz. Do their lifetimes increase?
  \item Turn off the lumped RLC boundaries of the flux bias lines; how does it affect the lifetimes of the qubits?
    \item Reduce the coupling between Q2 and R2 and between Q3 and R3 such that $|2\chi|$ is about 1 MHz. What are the qubit lifetimes?
    \item Add airbridges across the feedline and resonators (R2, RC, and R3) to connect the inner ground plane to the outer ground plane. Will Mode 2 and Mode 9 disappear? Follow the approach in~\cite{Dhillon_import}. 
\end{itemize}

\section{Conclusions}
In this paper, a step-by-step tutorial is created to demonstrate how to design a 3-qubit superconducting qubit chip. Essential theories are introduced before calculations and simulations in the design process. The source code and simulation projects are shared, and videos are recorded to assist learning. It covers feedline, resonator, transmon, transmon qubit, flux-bias lines, and qubit readout design. Finally, a 3-qubit chip is designed and demonstrated, and its drawbacks are discussed.
Extended exercises are proposed to further enhance the learning experience. After learning from this paper, students will be able to use HFSS to design and simulate their own superconducting qubit chips.

\section{Acknowledgment}
This work was supported by the National Science Foundation under Grant 2125906 and in part by AMDT Endowment Funding from the College of Engineering, San Jose State University.

\bibliographystyle{IEEEtran}
\bibliography{main}

\end{document}